# The Family of LML Detectors and the Family of LAS Detectors for Massive MIMO Communications

Yi Sun
Electrical Engineering Department
The City College of City University of New York, New York, NY 10031, USA
ysun@ccny.cuny.edu

**Abstract**

The family of local maximum likelihood (LML) detectors, including the global maximum likelihood (GML) detector, and the family of likelihood ascent search (LAS) detectors are akin to each other and possess common properties significant in both theory and practical multi-input multi-output (MIMO) communications. It is proved that a large MIMO channel possesses the LML characteristic, implying and predicting that a local search detector with likelihood ascent, like a wide-sense sequential LAS (WSLAS) detector, can approach the GML detection. By the replica method, the bit error rate (BER) of an LML detector in the large MIMO channel is obtained. The BER indicates that in the high signal-to-noise ratio (SNR) regime, both the LML and GML detectors achieve the AWGN channel performance when the channel load is as high as up to 1.5086 bits/dimension with an equal-energy distribution, and the channel load can be higher with an unequal-energy distribution. The analytical result is verified by simulation in the equal-energy distribution that the sequential LAS (SLAS) detector, a linear-complexity LML detector, can approach the BER of the NP-hard GML detector. The LML and LAS detectors in the two families are successfully applied to symbol detection in massive antenna MIMO communications and demonstrate the performance near the GML detection. This book chapter reviews the LML and LAS detectors in a unified framework. The focus is on their formulation, relationships, properties, and GML performance in BER and spectral efficiency in large MIMO channels.

## 1. Introduction

The family of likelihood ascent search (LAS) detectors originated from the modified Hopfield neural network (MHNN) [1-8]. When the original Hopfield neural network (HNN) [9] is applied to image restoration, a least squares problem with integer support, the error metric change is nonmonotonic – a phenomenon of instability [10]. To guarantee stability, the sequential MHNN [1] is proposed to update neurons one by one and check the error metric decrease in each step. The MHNN [2] with the parallel, partial parallel, and sequential updating modes is proposed for binary image restoration, which sets up the corresponding thresholds to ensure stability. The MHNN with the eliminating highest error (EHE) criterion is proposed in [3, 4] that updates one or multiple neurons with the largest error gradients in each step to converge to a better solution in fewer steps. A generalized updating rule (GUR) is proposed in [5, 6], by which an arbitrary set of neurons can be updated in each step while the error metric is guaranteed to decrease monotonically. In [7], several algorithms with particular sequences of updating modes are derived from the GUR and applied to gray image restoration and reconstruction. The EHE criterion and the fastest metric descent (FMD) criterion are used to design the sequence of updating modes so that the algorithms converge in fewer steps to better solutions with smaller errors and fewer strips in simulation. Analytical results [8] confirm that the EHE and FMD-based algorithms can achieve higher correct transition probabilities.

When applied to code division multiaccess (CDMA) communications, these algorithms under the GUR turn out to perform likelihood ascent search and form the family of LAS detectors [11-14]. Given any sequence of sets of updating candidates, the LAS detector searches out a sequence of bit vectors with monotonic likelihood ascent and, therefore, monotonically decreases the error probability of the bit vectors. In the LAS family, the wide-sense sequential LAS (WSLAS) detectors are particularly interesting that converge to a local maximum likelihood (LML) vector with neighborhood size one, the best detectors in the LAS family. The EHE and FMD criteria can be applied to enable the WSLAS detector to converge in fewer steps to a better solution [11]. By computationally efficiently updating the likelihood gradient in each step, all LAS detectors have a per-bit complexity linear in the number of transmitted bits. The concept of the LML detector is extended to any neighborhood size and forms the family of LML-$J$ detectors with any neighborhood size $J \geq 1$ [15-17]. The LML-$J$ detector can be realized by a likelihood ascent search in a neighborhood of size $J$ and the computation is reduced to the minimum by efficiently updating the likelihood gradient. The per-bit computational complexity of the LML-$J$ detector is in the order of $O(\sum_{i=1}^{J}\binom{K}{i})$ with $K$ transmitted bits. The WSLAS/LML-1 detectors belong to both the LAS and LML families. The bit error rate (BER) and asymptotic multi-bit efficiency (AME) of LAS detectors are analyzed in [19]. It is obtained that the LAS BER and AME are comparable to those of the global maximum likelihood (GML) detector [18] and demonstrate the superiority of LAS detectors over other detectors.

In large random spreading (LRS) CDMA channels, a random spectral-temporal spreading sequence spreads a bit, and both the number of bits and sequence dimension tend to infinity with their ratio kept constant. It is observed for the first time in simulation [20] that in a random spreading CDMA system, as the number of bits increases, the sequential LAS (SLAS) detector (a special WSLAS detector) monotonically decreases BER. When the number of bits reaches 200, the SLAS detector already approaches the single-bit BER under a channel load of 0.5 bits/s/Hz and signal-to-noise ratio (SNR) of 8 dB with an equal-energy distribution. It is essential that the LRS CDMA channels are proven to possess the "LML characteristic" [21, 22]; that is, the channel output retains the local topology of the input set of bit vectors. Therefore, "to achieve the GML detection one would, without the exhaustive search over the entire set, perform only an LML-1 detection." It is shown that the WSLAS/LML-1 detector achieves unit AME with channel load $< 1/2 - 1/(4 \ln 2)$ bits/dimension and, therefore, the single-bit BER in the high SNR regime. In simulations, for an equal-power distribution, the WSLAS/LML-1 detector with a linear per-bit complexity approaches the NP-hard GML BER in all SNR with channel load as high as 1.05 bits/dimension [21, 22]. By employing sparse spreading sequences where most chips are zero, the WSLAS/LML-1 detectors need only about 16 nonzero chips to achieve the same BER performance with fully spread sequences [23, 24]. The analytical results in [21, 22] are extended to the general large MIMO channel in [25]. The BER of the WSLAS and LML-$J$ detectors with fixed $J$ is obtained by using the powerful replica method. It is shown that the WSLAS/LML-1 detector consistently achieves one solution of the GML BER in all channel conditions. In a wide practical range of channel load and SNR, the WSLAS/LML-1 detector performs identically to the GML detector and achieves spectral efficiency as high as 1.5086 bits/dimension and the single-bit BER bound in the high SNR regime in an equal-energy distribution

When the LAS detector is applied to massive antenna MIMO systems in [26], it demonstrates excellent performance in low BER and linear per-bit complexity. The results triggered the extensive study of the family of LAS detectors and the development of low-complexity near-optimum detectors with local search [27, 28-58]. Various detectors are used as the initial detector of the LAS detector [26, 28-35]. The reactive tabu search is applied to avoid previously visited vectors in the future search [36-42]. The set of candidate vectors in each step is appropriately selected and reduced to improve BER performance and reduce computational complexity [43-46]. The LAS detector starts with multiple initial vectors and searches through several sequences of vectors, and then the best output vector is chosen [47-48]. The LAS detector and the LML-$J$ detector for $J = 2, 3$ are cascaded to improve the BER performance at the cost of increased computational complexity [28, 49]. The performance of the LAS detector is investigated when channel state information is imperfect [50-52]. The LAS detector is applied to joint symbol detection and channel estimation [53], decoding [54], and antenna selection [55]. The LAS detector invoked specific hardware design [56] and implementation on an FPGA chip [35]. The LAS detector is combined with the convolutional neural network [57] and deep learning [58]. The key feature of the family of LAS detectors is to search a sequence of bit vectors with a monotonic likelihood ascent and converge to a fixed point in a finite number of search steps. The key feature of the family of LML detectors is to search out an LML-$J$ vector with neighborhood size $J$. Hence, while there are various variations, many of the abovementioned local search detectors belong to the family of LAS detectors and/or the family of LML detectors.

It is not surprising that the LAS detector and other local search detectors in the literature are successful and approach the GML performance in the massive antenna MIMO channels [26-58]. The LRS CDMA channel [21, 22], the sparse LRS CDMA channel [23, 24], and the massive antenna MIMO channel [27] are particular instances of the general large MIMO channel [25]. These large MIMO channels have a common property: as the dimension of channel vectors tends to infinity, the crosscorrelation coefficient between two channel vectors converges to zero. This phenomenon is termed "channel hardening" for massive antenna MIMO channels [27]. It is because of this property that the general large MIMO channel possesses the "LML characteristic." Therefore, as prophesied in [21, 22], in a practical range of channel load and SNR, "there is typically no LML but the GML point. The likelihood function is sufficiently smooth. Thus, the linear-complex WSLAS detector (and other detectors) with local search can reach the GML point that is usually NP-hard to obtain. All the results are also applicable to the LML detectors with neighborhood sizes greater than one [15-17]."

As the number of mobile devices and the required spectral efficiency ever increase [59, 60], the next-generation mobile network will rely on massive MIMO communications to fulfill the ever-increasing demand, and then the LAS and LML detectors are expected to play a role. To this end, it is essential to understand the LML and LAS detectors and their properties. In this book chapter, we review the family of LML detectors and the family of LAS detectors in a unified framework. The focus is on their formulation, relationships, properties, and the GML BER and spectral efficiency in massive MIMO channels.

The rest of the book chapter is organized as follows. Section 2 presents the GML detector and the family of LML detectors. Section 3 addresses the family of LAS detectors and their performance in BER and AME. Section 4 studies the LML characteristic of massive MIMO channels and the analytical BER of the WSLAS and LML detectors by the replica method. Section 5 briefly discusses the application of LML and LAS families to the next generation multiaccess 5G/6G and beyond wireless networks. Conclusions are drawn in Section 6.

## 2. The Family of Local Maximum Likelihood (LML) Detectors

*2.1. The MIMO channel model*

Consider a general MIMO channel where $K$ complex symbols are transmitted through a complex Gaussian channel. Each symbol is equiprobably selected from an alphabet $\Omega_K$. The $k$th symbol $b_k$ is modulated by an $N$-dimensional complex channel vector $\boldsymbol{s}_k$ with unit length $\|\boldsymbol{s}_k\| = 1$. The energy of the $k$th symbol is $A_k^2$. The receiver receives an $N$-dimensional complex vector

$$\boldsymbol{r} = \mathbf{S}\mathbf{A}\boldsymbol{b} + \boldsymbol{m} \qquad (1)$$

where $\mathbf{S} = (\boldsymbol{s}_1, \dots, \boldsymbol{s}_K) \in C^{N \times K}$ is the channel matrix, $\mathbf{A} = \text{diag}(A_1, \dots, A_K)$ is the diagonal matrix of symbol amplitudes, $\boldsymbol{b} = (b_1, \dots, b_K)^T$ is the vector of symbols, and $\boldsymbol{m} \sim CN(\mathbf{0}, \sigma^2 \mathbf{I}_N)$ is the AWGN.

A $K$-dimensional sufficient statistic can be obtained by the matched filter (MF) $\mathbf{S}^H$ as

$$\boldsymbol{y} = \mathbf{S}^H \boldsymbol{r} = \mathbf{R}\mathbf{A}\boldsymbol{b} + \boldsymbol{n} \qquad (2)$$

where $\mathbf{R} = \mathbf{S}^H \mathbf{S}$ is the crosscorrelation matrix of channel vectors and $\boldsymbol{n} = \mathbf{S}^H \boldsymbol{m} \sim CN(\mathbf{0}, \sigma^2 \mathbf{R})$ with $H$ denoting the Hermitian transpose is the noise at the MF output. The probability density function of $\boldsymbol{y}$ given $\boldsymbol{b}, \mathbf{S}, \mathbf{A}$ is

$$p(\boldsymbol{y}|\boldsymbol{b}, \mathbf{S}) = \frac{1}{(\pi \sigma^2 |\mathbf{R}|)^K} \exp\left( -\frac{(\boldsymbol{y} - \mathbf{R}\mathbf{A}\boldsymbol{b})^H \mathbf{R}^{-1}(\boldsymbol{y} - \mathbf{R}\mathbf{A}\boldsymbol{b})}{\sigma^2} \right) \qquad (3)$$

where $\|\cdot\|$ is the $l_2$ norm. When $\mathbf{R}$ is not invertible, $\mathbf{R}^{-1}$ denotes its pseudoinverse. The posterior probability of $\boldsymbol{b}$ is

$$p(\boldsymbol{b}|\boldsymbol{y}, \mathbf{S}) = \frac{p(\boldsymbol{b})p(\boldsymbol{y}|\boldsymbol{b}, \mathbf{S})}{\sum_{\boldsymbol{a} \in \Omega_K} p(\boldsymbol{a}) p(\boldsymbol{y}|\boldsymbol{a}, \mathbf{S})}. \qquad (4)$$

We will use both $\boldsymbol{r}$ and $\boldsymbol{y}$ henceforth whenever one is more suitable.

The general MIMO channel model in Eq. (1) includes several typical MIMO communication systems. First, multiple users, each equipped with one or multiple transmit antennas, access a basestation (BS) that is equipped with $N$ receive antennas [27]. Second, in a CDMA system where $K$ users access a BS with one antenna, $\boldsymbol{s}_k$ represents the spreading sequence of the $k$th user and $N$ is the spreading factor [18, 19]. Third, by the multicode technique with extended symbol periods, each user can transmit multiple symbols with/without spread spectrum and/or time for single or multiple users [21, 22]. The codes can be long, short, and sparse [22-24]. Fourth, the multi-transmit antenna, multi-receive antenna, CDMA, and multicode techniques can be applied in combination, and each yields the same channel model in Eq. (1).

The GML detector, the family of LML detectors, and the family of LAS detectors can all be applied to any complex symbol constellation. As often considered in the literature, we shall subsequently focus on the BPSK signals transmitted through a real AWGN channel. Each symbol independently and equiprobably takes on $\{-1,1\}$ and then $\boldsymbol{b} \in \Omega_K \equiv \{-1,1\}^K$. The channel noise is $\boldsymbol{m} \sim N(\mathbf{0}, \sigma^2 \mathbf{I}_N)$. Accordingly, the probability density function of $\boldsymbol{r}$ conditioned on $(\boldsymbol{b}, \mathbf{S})$ is obtained in (3) by replacing $\sigma^2$ and $N$ with $2\sigma^2$ and $N/2$, respectively. $\mathbf{S}, \boldsymbol{b}$, and $\boldsymbol{m}$ are assumed to be mutually independent. The prior probability distribution of $\boldsymbol{b}$ is $p(\boldsymbol{b}) = 2^{-K}, \forall \boldsymbol{b} \in \Omega_K$. The average symbol energy is normalized to unit $K^{-1} \sum_{k=1}^{K} A_k^2 = 1$. The channel load is $\alpha = K/N$ in the unit of bits/dimension. In a system where the channel dimension is the product of time and bandwidth without spatial diversity, the channel load is in the unit of bits/s/Hz. Such a system can be the CDMA system where the channel vectors are code sequences with/without spread spectrum and/or time [21, 22].

## 2.2. The global maximum likelihood (GML) detector

A bit vector $\boldsymbol{b}$ can be demodulated from the received signal $\boldsymbol{r}$. With equiprobable bits, the probability density function $p(\boldsymbol{y}|\boldsymbol{b},\mathbf{S})$ is also a likelihood function of $\boldsymbol{b}$ with given the signal $\boldsymbol{y}$. Equivalently, a likelihood function can be defined from Eq. (3) as

$$f(\boldsymbol{y}|\boldsymbol{b}) = -\frac{1}{2}(\boldsymbol{y}-\mathbf{R}\mathbf{A}\boldsymbol{b})^H \mathbf{R}^{-1}(\boldsymbol{y}-\mathbf{R}\mathbf{A}\boldsymbol{b}). \tag{5}$$

The GML detector $\varphi^{\text{GML}}$ selects the bit vector that achieves the maximum likelihood among all possible vectors

$$\widehat{\boldsymbol{b}} = \arg\max_{\boldsymbol{b}\in\Omega_K} f(\boldsymbol{y}|\boldsymbol{b}), \quad \forall \boldsymbol{y}. \tag{6}$$

With equiprobable bits, the GML detector also achieves the global maximum posterior probability $\widehat{\boldsymbol{b}} = \arg\max_{\boldsymbol{b}\in\Omega_K} p(\boldsymbol{b}|\boldsymbol{y},\mathbf{S})$.

By maximizing the likelihood function globally over $\Omega_K$ for any $\boldsymbol{r}$, the GML detector minimizes the probability of error $P_e(\varphi) \equiv \Pr(\boldsymbol{b}^\varphi \neq \boldsymbol{b})$ among all detectors $\varphi$, and so is the optimum detector. On the other hand, the GML detector needs to compare the likelihood of all $2^K$ vectors in $\Omega_K$ and, therefore, is NP-hard in computation. The GML detector is usually infeasible in practical communication systems when $K$ is large.

## 2.3. Local maximum likelihood (LML) detectors

To trade off error performance for a low computational complexity, the LML detectors can be applied [15-17]. An LML detector depends on the neighborhood size of a bit vector. The neighborhood of $\boldsymbol{b}$ with neighborhood size $J$ is defined by the set of vectors that differ from $\boldsymbol{b}$ by at most $J$ bits

$$\Omega_J(\boldsymbol{b}) = \{\boldsymbol{a} \in \Omega_K | \, \|\boldsymbol{a}-\boldsymbol{b}\|_1/2 \leq J\}$$

where $\|\cdot\|_1$ is the $l_1$ norm.

Given $\boldsymbol{y}$, an LML detector $\varphi_J^{\text{LML}}$ with neighborhood size $J$ chooses a vector $\widehat{\boldsymbol{b}}$ that attains the maximum likelihood in its neighborhood of size $J$ [15-17]

$$\widehat{\boldsymbol{b}} = \arg\max_{\boldsymbol{b}\in\Omega_J(\widehat{\boldsymbol{b}})} f(\boldsymbol{y}|\boldsymbol{b}), \quad \forall \boldsymbol{y}. \tag{7}$$

With the equiprobable bits, the LML detector also achieves the local maximum posterior probability $\widehat{\boldsymbol{b}} = \arg\max_{\boldsymbol{b}\in\Omega_J(\widehat{\boldsymbol{b}})} p(\boldsymbol{b}|\boldsymbol{r},\mathbf{S})$.

In the particular case of $J = K$, the neighborhood of a vector is the entire set of all vectors $\Omega_J(\widehat{\boldsymbol{b}}) = \Omega_K$. The LML detector in Eq. (7) becomes the GML detector. Henceforth, for simplicity, LML-$J$ denotes LML with neighborhood size $J$.

An LML-$J$ vector for $J < K$ is a fixed point of Eq. (7), and its righthand side has a neighborhood depending on the vector $\widehat{\boldsymbol{b}}$ to be determined. In contrast, the GML (i.e., LML-$K$) vector in Eq. (6) has the neighborhood of the entire vector set independent of the vector to be determined.

## 2.4. GML and LML regions

The performance of the LML and GML detectors can be understood by the LML and GML vectors and their decision regions.

Given $\boldsymbol{y}$, an LML-$J$ vector solves Eq. (7) and the set of LML vectors with neighborhood size $J$ is defined as [15-17]

$$\Psi_J^{\text{LML}}(\boldsymbol{y}) = \{\boldsymbol{b} \in \Omega_K | f(\boldsymbol{y}|\boldsymbol{b}) \geq f(\boldsymbol{y}|\boldsymbol{a}), \forall \boldsymbol{a} \in \Omega_J\backslash\{\boldsymbol{b}\}\}. \tag{8}$$

In particular, the set of GML vectors is

$$\Psi^{\text{GML}}(\boldsymbol{y}) = \Psi_K^{\text{LML}}(\boldsymbol{y}) = \{\boldsymbol{b} \in \Omega_K | f(\boldsymbol{y}|\boldsymbol{b}) \geq f(\boldsymbol{y}|\boldsymbol{a}), \forall \boldsymbol{a} \in \Omega_K\backslash\{\boldsymbol{b}\}\}.$$

Depending on $\boldsymbol{y}$, there might be multiple vectors in $\Psi_J^{\text{LML}}(\boldsymbol{y})$ that solve Eq. (7). On the other hand, a unique GML vector solves Eq. (6), that is, $\Psi^{\text{GML}}(\boldsymbol{y}) = \{\boldsymbol{b}^{\text{GML}}(\boldsymbol{y})\}$ unless $\boldsymbol{y}$ is on the boundary of multiple decision regions, which occurs with probability zero. Hence, the GML vector is unique in general.

Since an LML vector with neighborhood size $J + 1$ is also an LML vector with neighborhood size $J$, the following relationships hold

$$\Psi^{\text{GML}}(\boldsymbol{y}) \subseteq \Psi_{K-1}^{\text{LML}}(\boldsymbol{y}) \subseteq \cdots \subseteq \Psi_2^{\text{LML}}(\boldsymbol{y}) \subseteq \Psi_1^{\text{LML}}(\boldsymbol{y}). \tag{9}$$

Hence, an LML-$J$ vector is also an LML-$M$ vector for $M < J$. The GML vector is an LML-$J$ vector for all $J < K$.

The relationships in Eq. (9) imply two properties of the family of LML detectors. First, it is evident that all the equalities in Eq. (9) hold with probability one (i.e., $\forall \boldsymbol{y}$ except the boundary of decision regions) iff in the trivial orthogonal channel where the channel vectors $\boldsymbol{s}_k$ are orthogonal. When there are multiple LML-$J$ vectors in $\Psi_J^{\text{LML}}(\boldsymbol{y})$, an LML-$J$ detector can output any LML-$J$ vector and, therefore, the LML detector $\varphi_J^{\text{LML}}$ is not unique for $J < K$. Second, when there are multiple LML-$J$ vectors, though all the LML-$J$ vectors in $\Psi_J^{\text{LML}}(\boldsymbol{y})$ solve Eq. (7), some of the LML-$J$ vectors have a higher likelihood than the others. For $\boldsymbol{a}, \boldsymbol{b} \in \Psi_J^{\text{LML}}(\boldsymbol{y})$, if $\boldsymbol{a} \in \Psi_{J+1}^{\text{LML}}(\boldsymbol{y})$ but $\boldsymbol{b} \notin \Psi_{J+1}^{\text{LML}}(\boldsymbol{y})$, then the likelihood of $\boldsymbol{a}$ is higher than $\boldsymbol{b}$. In other words, in addition to the LML-$(J + 1)$ vectors in $\Psi_{J+1}^{\text{LML}}(\boldsymbol{y})$, $\Psi_J^{\text{LML}}(\boldsymbol{y})$ contains some additional LML-$J$ vectors whose likelihoods are lower than the likelihoods of all LML-$(J + 1)$ vectors in $\Psi_{J+1}^{\text{LML}}(\boldsymbol{y})$. An LML-$J$ detector $\varphi_J^{\text{LML}}$ selects one vector in $\Psi_J^{\text{LML}}(\boldsymbol{y})$ and achieves a probability of error not less than an LML-$(J + 1)$ detector. Hence, the following relationships hold

$$P_e(\varphi^{\text{GML}}) \leq P_e(\varphi_{K-1}^{\text{LML}}) \leq \cdots \leq P_e(\varphi_2^{\text{LML}}) \leq P_e(\varphi_1^{\text{LML}}). \tag{10}$$

Note that in Eq. (10), the equality in $P_e(\varphi_{J+1}^{\text{LML}}) \leq P_e(\varphi_J^{\text{LML}})$ holds iff $\varphi_J^{\text{LML}}$ chooses only the vectors belonging to $\Psi_{J+1}^{\text{LML}}(\boldsymbol{y})$ among all vectors in $\Psi_J^{\text{LML}}(\boldsymbol{y})$; but this means that $\varphi_J^{\text{LML}}$ is an LML-$(J + 1)$ detector. Hence, all the inequalities in Eq. (10) hold in general, except for the trial orthogonal channel.

The LML-$J$ region of a bit vector $\boldsymbol{b} \in \Omega_K$ is defined as the region of $\boldsymbol{y} \in \mathbb{R}^K$ such that $\boldsymbol{b}$ is an LML-$J$ vector [15-17]

$$V_J^{\text{LML}}(\boldsymbol{b}) = \{\boldsymbol{y} \in \mathbb{R}^K | f(\boldsymbol{y}|\boldsymbol{b}) \geq f(\boldsymbol{y}|\boldsymbol{a}), \forall \boldsymbol{a} \in \Omega_J \setminus \{\boldsymbol{b}\}\}. \tag{11}$$

In particular, the GML region is the LML-$K$ region

$$V^{\text{GML}}(\boldsymbol{b}) = V_K^{\text{LML}}(\boldsymbol{b}) = \{\boldsymbol{y} \in \mathbb{R}^K | f(\boldsymbol{y}|\boldsymbol{b}) \geq f(\boldsymbol{y}|\boldsymbol{a}), \forall \boldsymbol{a} \in \Omega_K \setminus \{\boldsymbol{b}\}\}.$$

For any $\boldsymbol{y} \in V_{J+1}^{\text{LML}}(\boldsymbol{b})$, $\boldsymbol{b}$ is an LML-$(J + 1)$ vector. Then $\boldsymbol{b}$ is also an LML-$J$ vector. Consequently, $\boldsymbol{y} \in V_J^{\text{LML}}(\boldsymbol{b})$. Hence, the following relationships hold

$$V^{\text{GML}}(\boldsymbol{b}) \subseteq V_{K-1}^{\text{LML}}(\boldsymbol{b}) \subseteq \cdots \subseteq V_2^{\text{LML}}(\boldsymbol{b}) \subseteq V_1^{\text{LML}}(\boldsymbol{b}). \tag{12}$$

In the region of $V_J^{\text{LML}}(\boldsymbol{b}) \setminus V_{J+1}^{\text{LML}}(\boldsymbol{b})$, $\boldsymbol{b}$ is an LML-$J$ vector but is not an LML-$(J + 1)$ vector. In other words, in addition to $V_{J+1}^{\text{LML}}(\boldsymbol{b})$, $V_J^{\text{LML}}(\boldsymbol{b})$ contains an extra region where $\boldsymbol{b}$ is an LML-$J$ vector and $\boldsymbol{b}$ has a likelihood not greater than the other LML-$(J + 1)$ vectors. This also implies that Eq. (10) is true. Hence, as the neighborhood size $J$ increases, the error probability of an LML-$J$ detector monotonically decreases.

Figure 1 (a) in the next section shows the LML and GML regions with neighborhood size $J = 1$ and $J = 2$, respectively, in a two-bit channel with $R_{12} = \rho > 0$. A more detailed explanation of the LML and GML regions is presented in the paragraph below Figure 1.

Given $K \geq 3$, as the neighborhood size $J$ increases, the region where multiple LML-$J$ vectors coexist shrinks, and then the BER performance of an LML-$J$ detector improves. In general, the GML detector performs substantially better than an LML-$J$ detector with a small $J$ in a small- or median-size system [15-17]. However, as presented in Section 4, in a large MIMO system where both $K$ and $N$ tend to infinity at the same rate, an LML-*J* detector with a small *J* can perform equally well as the GML detector in a wide range of channel load and SNR [22, 25].

*2.5. The family of LMLAS detector*

The family of local-maximum-likelihood likelihood-ascent-search (LMLAS) detectors [15-17] can achieve the LML-*J* detection for $J \in \{1, \ldots, K\}$. For simplicity, LMLAS-*J* means LMLAS with neighborhood size *J*.

The family of LMLAS detectors is designed in principle upon Eq. (7). Since both sides of Eq. (7) depend on $\hat{\boldsymbol{b}}$, the LML-*J* vector $\hat{\boldsymbol{b}}$ is a fixed point that solves Eq. (7). The definition of the LML-*J* detector in Eq. (7) suggests that an LML-*J* vector can be obtained by an iterative search [15-17]. Given $\boldsymbol{y}$, starting with an initial vector $\boldsymbol{b}(0)$, search for a higher-likelihood vector $\boldsymbol{b}(1)$ in $\Omega_J(\boldsymbol{b}(0))$, the neighborhood of $\boldsymbol{b}(0)$ with neighborhood size *J*; if $\boldsymbol{b}(1)$ does not attain the maximum likelihood in its neighborhood $\Omega_J(\boldsymbol{b}(1))$, then search for a higher-likelihood vector $\boldsymbol{b}(2) \in \Omega_J(\boldsymbol{b}(1))$; repeat the search until an LML-*J* vector $\hat{\boldsymbol{b}}$ is obtained, which achieves the maximum likelihood in its own neighborhood $\Omega_J(\hat{\boldsymbol{b}})$. To remove redundant computations, the search is guided by the likelihood gradient of the current vector, and the gradient is updated computationally efficiently by Eq. (18) in the next section.

*2.6. Computational complexity*

An LMLAS detector can computationally efficiently perform the likelihood ascent search [15-17]. To avoid redundant computation, the LMLAS detectors employ the gradient of the likelihood function $f(\boldsymbol{y}|\boldsymbol{b})$. The likelihood of a vector can be compared with the likelihood of the current vector in terms of the gradient of likelihood function at the current vector. Once a vector with a higher likelihood is accepted as the current vector, the gradient can be updated by parallel vector additions of Eq. (18). Since the likelihood comparison and the gradient update are two core computations and both can be computationally efficiently performed, the family of LMLAS detectors achieves a high efficiency in computation. To determine an LML-*J* vector, it needs to compare the likelihood of the current vector with other vectors in the neighborhood of size *J*. Thus, the per-bit computational complexity of an LMLAS-*J* detector is in the order of [15-17]

$$C_J^{\text{LML}} = O\left(\sum_{j=1}^{J} \binom{K}{j}\right), \tag{13}$$

and, therefore,

$$C^{GML} = C_K^{\text{LML}} > \cdots > C_2^{\text{LML}} > C_1^{\text{LML}}. \tag{14}$$

In particular, the LMLAS-1 detector achieves the lowest computational complexity $C_1^{\text{LML}} = O(K)$ linear in *K*. The LMLAS-*K* detector (i.e. the GML detector) is the most complex with a computational complexity $C^{GML} = C_K^{\text{LML}} = O(2^K)$ exponential in *K*.

## 3. The Family of Likelihood Ascent Search (LAS) Detectors

By choosing a different neighborhood size *J*, the family of LMLAS detectors can achieve a different tradeoff of computational complexity and error performance. An LMLAS-1 detector achieves the lowest computational complexity and the worst error performance. To enable further reduction of computational complexity, the family of LAS detectors can be applied [11-14, 19-24].

*3.1. Criteria for the updating rule*

The idea of designing the family of LAS detectors is to search a sequence of bit vectors $\boldsymbol{b}(n)$ such that the likelihood of $\boldsymbol{b}(n)$ monotonically increases with step *n* until reaching a fixed point. The bits scheduled to update in a step are called update candidates. Updating a bit is to check the flip condition of the bit, but it may or may not result in a bit flip according to the updating rule.

Suppose $L(n) \subseteq \{1, \ldots, K\}$ is the index set of update candidates in the $n$th step. The gradient $\nabla f(\mathbf{b})$ of the likelihood function with respect to $\mathbf{b}$ evaluated at $\mathbf{b}(n)$ equals

$$\mathbf{g}(n) = -\mathbf{H}\mathbf{b}(n) + \mathbf{A}\mathbf{y} \quad (15)$$

with $\mathbf{H} = \mathbf{A}\mathbf{R}\mathbf{A}$ whose diagonal elements are the symbol energies $H_{kk} = A_k^2$ as $R_{kk} = 1$ for any $k$. $\mathbf{g}(n)$ is simply equal to $\mathbf{A}$ times the difference between the MF output and the signal reconstructed by $\mathbf{b}(n)$, i.e., $\mathbf{g}(n) = \mathbf{A}[-\mathbf{R}\mathbf{A}\mathbf{b}(n) + \mathbf{y}]$. The $k$th component of the likelihood gradient is

$$g_k(n) = -\sum_{i=1}^{K} H_{ki} b_i(n) + A_k y_k. \quad (16)$$

If the bits in $L_p(n) \subseteq L(n)$ are flipped in the step $n$, then the new bit vector is

$$\mathbf{b}(n+1) = \mathbf{b}(n) - 2 \sum_{k \in L_p(n)} b_k(n) \mathbf{e}_k \quad (17)$$

where the $k$th element of $\mathbf{e}_k$ is equal to one, and others are zero. The likelihood gradient in the next step can be efficiently updated by

$$\mathbf{g}(n+1) = \mathbf{g}(n) + 2 \sum_{k \in L_p(n)} b_k(n) \mathbf{h}_k \quad (18)$$

where $\mathbf{h}_k$ is the $k$th column vector of $\mathbf{H}$. The likelihood change $\Delta f(n) = f[\mathbf{b}(n+1)] - f[\mathbf{b}(n)]$ can be calculated in terms of $\mathbf{g}(n)$ as

$$\Delta f(n) = \Delta \mathbf{b}^T(n) \left[ \mathbf{g}(n) + \tfrac{1}{2} \mathbf{z}(n) \right] \quad (19)$$

with $\Delta \mathbf{b}(n) = \mathbf{b}(n+1) - \mathbf{b}(n)$ and $\mathbf{z}(n) = -\mathbf{H}\Delta\mathbf{b}(n)$.

The LAS detector is developed based on the criteria that (i) the updating rule is computationally efficient, (ii) the new bit vector $\mathbf{b}(n+1)$ must have a higher likelihood than $\mathbf{b}(n)$ if $\mathbf{b}(n+1) \neq \mathbf{b}(n)$, and (iii) under the same framework of an updating rule, $\mathbf{b}(n+1)$ has the highest likelihood. It is shown that a LAS detector satisfies these criteria [11-14, 19].

### 3.2. The LAS detectors

The following generalized LAS detector applies to all possible sequences of candidate sets $L(n)$, $n \geq 0$ and defines the family of LAS detectors.

*LAS detector*: Given $L(n) \subseteq \{1, \ldots, K\}$ for all $n \geq 0$ and an initial vector $\mathbf{b}(0) \in \Omega_K$. At step $n$ all the bits for $k \in L(n)$ are updated by

$$b_k(n+1) = \begin{cases} +1, & \text{if } b_k(n) = -1 \text{ and } g_k(n) > t_k(n), \\ -1, & \text{if } b_k(n) = +1 \text{ and } g_k(n) < -t_k(n), \\ b_k(n), & \text{otherwise,} \end{cases} \quad (20)$$

where the $k$th threshold is determined by the elements of $\mathbf{H}$ as

$$t_k(n) = \sum_{j \in L(n)} |h_{kj}|, \quad (21)$$

all the bits for $k \notin L(n)$ remain unchanged $b_k(n+1) = b_k(n)$, and then $\mathbf{g}(n+1)$ is updated by Eq. (18) in which $L_p(n)$ is the index set of flipped bits in Eq. (20). $\mathbf{b}^*$ is the final demodulated vector if $\mathbf{b}(n) = \mathbf{b}^*$, $\forall n \geq n^*$ with some $n^* \geq 0$.

In the LAS update rule of Eq. (20), if the thresholds are too low, the iteration will enter a limit cycle. On the other hand, if the thresholds are too high, after updating the likelihood of $\mathbf{b}(n+1)$ is not sufficiently high, and the search will stop at a bit vector with a low likelihood. The following theorem indicates that the thresholds in the LAS detector are optimum [12-14].

*Theorem 1*: For any $L(n) \subseteq \{1, \ldots, K\}$ and **H**, the thresholds $t_k(n)$ in Eq. (21) are necessary and sufficient for the LAS detector to increase the likelihood of a nonzero update $\boldsymbol{b}(n+1) \neq \boldsymbol{b}(n)$.

In practice, $L(n)$ for $n \geq 0$ can be scheduled so that all the bits are finally periodically updated. If no bit is flipped in a period, the LAS detector has reached a fixed point $\boldsymbol{b}^*$ and shall terminate. It is assumed that every bit is updated once more without a flip before reaching a fixed point. The sequence of $L(n)$ is deterministic in most cases, though it can be random. The LAS detector is step-invariant if $L(n_1) = L(n_2)$ for all $\boldsymbol{b}(n_1) = \boldsymbol{b}(n_2)$, $n_1 \neq n_2$.

Specifying a sequence of $L(n)$ for $n \geq 0$, one determines a particular LAS detector. One of the most straightforward sequences is to update one bit in each step, which produces a sequential LAS (SLAS) detector with the lowest and step-invariant thresholds

$$t_k = A_k^2. \tag{22}$$

The SLAS detector can update the bits in a circular or random order.

All the SLAS detectors belong to the larger set of wide-sense SLAS (WSLAS) detectors that set $|L(n)| = 1$, $\forall n \geq n'$ with some $n' \geq 0$. The WLSAS detector eventually updates one bit in each step and works in the SLAS mode. All the WSLAS detectors, including the SLAS detector, converge to an LML-1 vector and, therefore, are the LML-1 detectors [11-14].

Another simplest sequence is $L(n) = \{1, \ldots, K\}$, $\forall n \geq 0$, which yields the parallel LAS (PLAS) detector with the highest and step-invariant thresholds

$$t_k = \sum_{j=1}^{K} |h_{kj}|. \tag{23}$$

The PLAS detector updates all $K$ bits in each step.

The PLAS and the SLAS detectors are particular instances of the group-parallel LAS (GPLAS) detectors that update bits group by group. If $\zeta$ is a collection of subsets that partitions $\{1, \ldots K\}$, a GPLAS detector has $L(n) \in \zeta$ with thresholds

$$t_k = \sum_{j \in L} |h_{kj}|, \quad \forall k \in L \in \zeta. \tag{24}$$

More complicated sequences of $L(n)$, $n \geq 0$ can be specified, such as those WSLAS detectors based on the EHE and FMD criteria [11] to converge to a fixed point with a higher likelihood in fewer steps.

An initial vector $\boldsymbol{b}(0)$ with a lower error probability can make the LAS detector converge faster to a fixed point of lower error probability. To reduce dependency on the initial computational cost, a random vector in $\Omega_K$ and the MF detector output $\boldsymbol{b}(0) = \text{sgn}(\boldsymbol{y})$ can be employed as the initial vector.

The updating condition in Eq. (20) can be concisely written as $b_k(n)g_k(n) < -t_k(n)$ but it is elaborately written in the form of Eq. (20) for convenience of hardware implementation. Though Eq. (20) can also be rewritten without the use of the likelihood gradient $\boldsymbol{g}(n)$, there are good reasons to use it. First, the efficient updating of $\boldsymbol{g}(n)$ in Eq. (18) is the key to reducing redundant computations. Second, Eq. (18) can be implemented on hardware suitable for fast parallel computation like a multilayer perceptron. Third, searching the next vector along $\boldsymbol{g}(n)$ with likelihood ascent is the motivation of LAS design, which identifies the LAS detectors from the other detectors motivated by different criteria [64-67].

### 3.3. Monotonic likelihood ascent and stability

An iterative algorithm that searches out a sequence of bit vectors is stable if it monotonically increases an upper-bounded Lyapunov function, thus ensuring the convergence to a fixed point in a finite number of iterations. Otherwise, it enters a limit cycle on which the Lyapunov function is not monotonic. The likelihood function $f(\mathbf{y}|\mathbf{b})$ in Eq. (5) is a Lyapunov function of the LAS detector. The following theorem is proved [11-14, 19].

*Theorem 2*: Consider a LAS detector with $L(n), n \geq 0$ that generates a sequence of bit vectors $\mathbf{b}(n), n \geq 0$. (i) For any $\mathbf{y}$, $f[\mathbf{y}|\mathbf{b}(n+1)] \geq f[\mathbf{y}|\mathbf{b}(n)], \forall n \geq 0$ with equality iff $\mathbf{b}(n+1) = \mathbf{b}(n)$; (ii) $\mathbf{b}(n)$ converges to a fixed point $\mathbf{b}^*$, i.e., $\mathbf{b}(n) = \mathbf{b}^*$, $\forall n \geq n^*$ with a finite number of steps $n^* \geq 0$; (iii) $P_e[\mathbf{b}(n+1)] \leq P_e[\mathbf{b}(n)]$ with equality iff $\Pr[\mathbf{b}(n+1) = \mathbf{b}(n)] = 1$; (iv) $P_e(\mathbf{b}^*) \leq P_e[\mathbf{b}(0)]$ with equality iff $\mathbf{b}(0)$ is a fixed point of the LAS detector with probability one, and for the WSLAS detector with equality iff $\mathbf{b}(0)$ is an LML-1 vector with probability one.

The theorem indicates that a LAS detector monotonically increases the likelihood, is stable, and monotonically reduces the probability of error for the sequence of searched vectors. Unless the initial detector is a fixed point with probability one, the LAS detector reduces the error probability of the initial detector.

### 3.4. Tradeoff between performance and computational complexity

The LAS detector can achieve a tradeoff between the error performance and the computational complexity by a proper design of the sequence of $L(n)$ for $n \geq 0$, the bits to be updated in each step. The following theorem is obtained [12]. The first inequality is obtained from the updating rule in Eq. (20) and the second inequality is further obtained from the fact of $f(\mathbf{y}|\mathbf{b}) = -\frac{1}{2}\mathbf{g}^H\mathbf{H}^{-1}\mathbf{g}$.

*Theorem 3*: Suppose a LAS detector converges to a fixed point $\mathbf{b}^*$ after $n^*$ steps, $\bigcup_{n \geq n^*} L(n) = \{1, \dots, K\}$, and $\mathbf{b}^*$ is not bounded. At $\mathbf{b}^*$ the gradient of the likelihood function is upper bound by

$$\|\mathbf{g}^*\|_1 \leq \sum_{k=1}^{K} \min_{n \geq n^*, k \in L(n)} \{t_k(n)\} \tag{25}$$

and the likelihood is lower bounded by

$$f(\mathbf{y}|\mathbf{b}^*) \geq -\frac{1}{2}\|\mathbf{H}^{-1}\| \sum_{k=1}^{K} \min_{n \geq n^*, k \in L(n)} \{t_k^2(n)\}. \tag{26}$$

Theorem 3 implies that the fewer bits that are updated at each step after reaching a fixed point (i.e., $|L(n)|$ for $n \geq n^*$ is smaller), the lower the thresholds, and then the lower the upper bounds of $\|\mathbf{g}^*\|_1$ and the higher the lower bound on the likelihood $f(\mathbf{y}|\mathbf{b}^*)$. In other words, a smaller $|L(n)|$ for $n \geq n^*$ results in a higher likelihood of the demodulated bit vector $\mathbf{b}^*$.

On the other hand, the more bits to be updated at a step, then the larger the search region in $\Omega_K$ (i.e., a possible larger $\|\mathbf{b}(n+1) - \mathbf{b}(n)\|_1$), the higher the thresholds, the fewer steps to converge to a fixed point. A convergence in fewer steps is particularly preferred when the LAS detector is implemented on hardware like a multilayer perceptron suitable for parallel computations where the computational time for demodulating a bit vector depends on the number of steps instead of the total number of additions and multiplications.

Hence, to trade off the error performance and the computational time, a better design of $L(n)$ for $n \geq 0$ is to update more bits with a large $|L(n)|$ in a few initial steps to achieve fast convergence and then gradually decrease the number of updated bits with a smaller $|L(n)|$, say $|L(n)| = 1$, to achieve the highest likelihood of $\mathbf{b}^*$ – an LML-1 vector, which is the criterion in the design of $L(n), n \geq 0$ for a WSLAS detector [11-15, 19].

The updating rule Eq. (20) with the gradient update Eq. (18) is the core cost of computational complexity and is the key to the linear complexity of the LAS detector. The total number of bit flips from $\mathbf{b}(0)$ to $\mathbf{b}^*$ equals $M = \sum_{n=0}^{n^*}|L_p(n)|$, which depends on $L(n), \mathbf{b}(0)$, and $\mathbf{y}$, and is random. Define the bit flip rate (BFR) as the average number of flips per bit $c = E(M)/K$. Computational complexity can be defined as the average number of additions per bit. Each bit flip results in $K$ additions in Eq. (18). The computational complexity equals $E(KM)/K = E(M) = cK$ linear in $K$. The BFR $c$ is about 0.5, as analyzed in [12-14] and less than 0.5, as observed in all simulations in various conditions, part of which is reported in [11, 21-24]. With the implementation of hardware suitable for vector additions, the computational complexity is further reduced by a factor of $K$.

### 3.5. Fixed-point region

The error performance of the LAS detector is determined by its output fixed point. Given $L(n)$ for $n \geq 0$, in the $\mathbf{y}$ space there is a region of $\mathbf{b}$ associated with the initial detector $\mathbf{b}(0)$ where the LAS detector converges from $\mathbf{b}(0)$ to $\mathbf{b}$. To determine the effect of a particular sequence $L(n)$ on error performance, define the fixed-point region of $\mathbf{b}$ by

$$V^{\text{LAS}}(\mathbf{b}) \equiv \{\mathbf{y} \in \mathbb{R}^K | \text{Exist } \mathbf{b}(0) \text{ s.t. the LAS detector converges to } \mathbf{b}\},$$

which is the union of such regions associated with all initial detectors. Conversely, given $\mathbf{y}$, define the set of fixed points

$$\Psi^{\text{LAS}}(\mathbf{y}) \equiv \{\mathbf{b} \in \Omega_K | \text{Exist } \mathbf{b}(0), L(n), n \geq 0 \text{ s.t. the LAS detector converges to } \mathbf{b}\}.$$

The following proposition is obtained [19].

*Proposition 1*: For any $\mathbf{y}$, denote by $\Lambda(\mathbf{y}, \mathbf{b}) = \{\mathbf{b}(0) \in \Omega_K | \text{LAS: } \mathbf{y}, \mathbf{b}(0) \to \mathbf{b}\}$ the set of initial bit vectors from which the LAS detector converges to **b**. Let $t_k^* = \max_{\mathbf{b}(0) \in \Lambda(\mathbf{y},\mathbf{b})} \min_{n \geq n_k^*[\mathbf{b}(0)], k \in L(n)} t_k(n)$ where $n_k^*[\mathbf{b}(0)]$ is the last flip step of the $k$th bit with the initial $\mathbf{b}(0)$. Then

$$V^{\text{LAS}}(\mathbf{b}) = \{\mathbf{y} \in \mathbb{R}^K | \mathbf{b} \otimes (\mathbf{Ay} - \mathbf{Hb}) \geq -\mathbf{t}^*\} \tag{27}$$

and conversely

$$\Psi^{\text{LAS}}(\mathbf{y}) = \{\mathbf{b} \in \Omega_K | \mathbf{b} \otimes (\mathbf{Ay} - \mathbf{Hb}) \geq -\mathbf{t}^*\} \tag{28}$$

where $\mathbf{t}^* = (t_1^*, \dots, t_K^*)^T$, and multiplication $\otimes$ and inequality $\geq$ are elementwise.

The SLAS, PLAS, and GPLAS detectors all have step-invariant thresholds and, therefore, $t_k^* = t_k$ where $t_k$ is given accordingly in Eqs. (22)-(24).

A WSLAS detector has the same fixed-point region of the SLAS detector as

$$V^{\text{WSLAS}}(\mathbf{b}) = \{\mathbf{y} \in \mathbb{R}^K | \mathbf{b} \otimes [\mathbf{y} - (\mathbf{R} - \mathbf{I})\mathbf{Ab}] \geq \mathbf{0}\}, \tag{29}$$

which is equal to $V_1^{\text{LML}}(\mathbf{b})$, the LML-1 region in Eq. (11). Hence, all the WSLAS detectors are the LML-1 detectors.

The fixed-point region by Eq. (27) can be applied to obtain the BER upper bound [19]. Moreover, Proposition 1 qualitatively indicates the characteristics of the fixed-point region and the relationship between the error performance and the thresholds. First, any bit vector $\mathbf{b}$ can be a fixed point with a nonzero probability. Second, given $\mathbf{y}$, the GML decision $\mathbf{b}^{\text{GML}}(\mathbf{y})$ is unique, and the GML decision regions $V^{\text{GML}}(\mathbf{b})$ for different $\mathbf{b}$'s do not overlap. In contrast, like all the LML-$J$ detectors for $J < K$, a LAS detector may have a set of multiple fixed points $\Psi^{\text{LAS}}(\mathbf{y})$, and the fixed-point regions $V^{\text{LAS}}(\mathbf{b})$ for different $\mathbf{b}$'s may overlap. In the overlapped fixed-point region, one of the fixed points is taken as the demodulated vector depending on the initial vector $\mathbf{b}(0)$ and the sequence of $L(n)$ for $n \geq 0$. Third, as the thresholds increase, the fixed-point region expands, the overlapped region expands, and the number of fixed points in the expanded region increases. The error probability increases since the increased fixed points have a lower likelihood. On the other hand, increasing the number of low-likelihood fixed points makes it easier to reach a fixed point, thus decreasing the computational complexity.

The LML-1, WSLAS, LAS, and PLAS detectors have the following duality relationships [19].

(i) For any $\mathbf{b}$, Proposition 1 implies that

$$V_1^{\text{LML}}(\mathbf{b}) = V^{\text{WSLAS}}(\mathbf{b}) \subseteq V^{\text{LAS}}(\mathbf{b}) \subseteq V^{\text{PLAS}}(\mathbf{b}). \tag{30}$$

All the equalities in Eq. (30) hold iff $\mathbf{R} = \mathbf{I}$ where all the GML, LML-$J$, and LAS detectors are collapsed to the $K$ parallel single-bit MF detectors.

(ii) For any $\mathbf{y} \in \mathbb{R}^K$, Proposition 1 implies that

$$\Psi_1^{\text{LML}}(\boldsymbol{y}) = \Psi^{\text{WSLAS}}(\boldsymbol{y}) \subseteq \Psi^{\text{LAS}}(\boldsymbol{y}) \subseteq \Psi^{\text{PLAS}}(\boldsymbol{y}). \tag{31}$$

The equality in Eq. (31) can be true for some $\boldsymbol{y}$ even when $\mathbf{R} \neq \mathbf{I}$.

The relationship $V^{\varphi}(\boldsymbol{b}) \subseteq V^{\phi}(\boldsymbol{b})$ in Eq. (30) means that the fixed-point region of detector $\varphi$ is inside that of $\phi$, which implies the relationship $\Psi^{\varphi}(\boldsymbol{y}) \subseteq \Psi^{\phi}(\boldsymbol{y})$ in Eq. (31) so that a solution of $\varphi$ is also a solution of $\phi$. $\phi$ performs worse than $\varphi$ only in the region $V^{\phi}(\boldsymbol{b}) \setminus V^{\varphi}(\boldsymbol{b})$. The error probabilities of these special LAS detectors have the following relationship

$$P_e(\varphi_1^{\text{LML}}) \cong P_e(\varphi^{\text{WSLAS}}) < P_e(\varphi^{\text{LAS}}) < P_e(\varphi^{\text{PLAS}}). \tag{32}$$

To understand the LML region and the LAS fixed point region, Figure 1 (a) shows the LML-1 and GML (i.e., LML-2) regions, respectively, in a two-bit channel with $R_{12} = \rho > 0$. The four dots are the noise-free MF-output signals $\mathbf{RA}\boldsymbol{b}$ for $\boldsymbol{b} \in \{-1, +1\}^2$. The LML-1 region of each vector $\boldsymbol{b} \in \{-1, +1\}^2$ where $\boldsymbol{b}$ is an LML-1 vector is bounded by two lines parallel to the axes. For instance, the LML-1 region of $(+1, -1)$ is bounded by the two lines $y_1 = -\rho A_2$ and $y_2 = \rho A_1$.

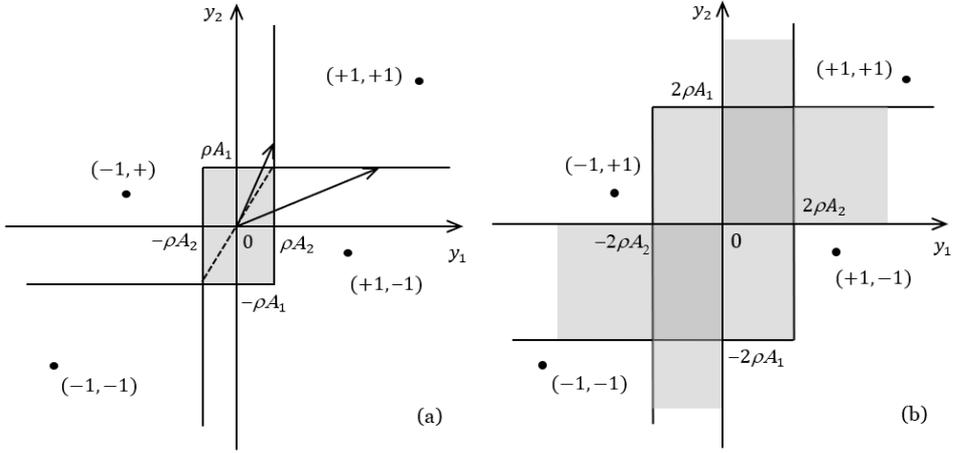

**Figure 1.** The $\boldsymbol{y}$ space for the two-bit channel with $\rho = 0.4$, $A_1 = 1$ and $A_2 = 0.6$. (a) The fixed-point regions (or LML-1 regions) of the WSLAS detectors. (b) The fixed-point regions of the PLAS detector.

The LML-1 regions of $(+1, +1)$ and $(-1, -1)$ are identical to the GML regions where an LML-1 detector performs identically to the GML detector. However, the LML-1 regions of $(+1, -1)$ and $(-1, +1)$ are overlapped in the shaded region where both vectors are LML-1 vectors and one of them is the GML vector depending on $\boldsymbol{y}$. The dotted line across the origin is the boundary of the GML regions of $(+1, -1)$ and $(-1, +1)$. In the shaded region, the vector $(-1, +1)$ is the GML vector above the dashed line, and the vector $(+1, -1)$ is the GML vector below the dashed line. There are infinitely many LML-1 detectors due to the infinitely many $\boldsymbol{y}$ each having two choices on $(+1, -1)$ and $(-1, +1)$. In the shaded region, an LML-1 detector will choose either $(+1, -1)$ or $(-1, +1)$ as the demodulated vector. In contrast, the GML detector will choose only the GML vector as the demodulated vector. Hence, only in the shaded region, an LML-1 detector performs worse than the GML detector.

As an LML-1 detector, a WSLAS detector converges to either $(+1, -1)$ or $(-1, +1)$ depending on the sequence of $L(n)$ for $n \geq 0$ and the initial vector $\boldsymbol{b}(0)$. The BER upper bound in the next section applies to all the LML-1/WSLAS detectors and thus is for the worst LML-1/WLAS detector that always chooses the low-likelihood vector in the overlapped region. In Figure 1 (b) of the PLAS fixed-point regions, the lightly shaded regions have two fixed points, and the deeply shaded regions have three fixed points. With higher thresholds, the PLAS detector has broader fixed-point regions and performs worse than a WSLAS detector.

By Eqs. (12) and (30), an LML-$J$ region with any neighborhood size $J$ is inside the fixed-point region of any LAS detector; and by Eqs. (9) and (31), an LML-$J$ vector with any neighborhood size is a fixed point of any LAS detector. Hence, a LAS detector does not change an initial vector that is an output of an LML-$J$ detector for any $J$. In contrast, since their decision region is not inside a fixed-point region of the LAS detector, none of the linear detectors (e.g., MF, zero-forcing (ZF), decorrelating, and MMSE detectors), SIC, PIC, DDF, and MMSE-DF detectors [1] can be a fixed point of any LAS detector with probability one. By Theorem 2, any LAS detector can reduce the error probabilities of these detectors, and a WSLAS detector can reduce them to local minima with neighborhood size $J = 1$.

The LML-1, WSLAS, SLAS, LAS, and PLAS detectors have the following relationship of computational complexity

$$C_1^{\text{LML}} \cong C^{\text{SLAS}} > C^{\text{WSLAS}} > C^{\text{LAS}} > C^{\text{PLAS}}. \tag{33}$$

In summary, the GML, LML-$J$ for $J = K-1, \ldots, 2, 1$, WSLAS, LAS, and PLAS detectors have the relationships of decision regions in Eqs. (12) and (30), sets of demodulated vectors in Eqs. (9) and (31), error probabilities in Eqs. (10) and (32), and computational complexities in Eqs. (14) and (33). Their relationships in error probability and computational complexity are shown in Figure 2. All these relationships imply that all the detectors in the LML and LAS families are akin to each other and are in the same class of detectors.

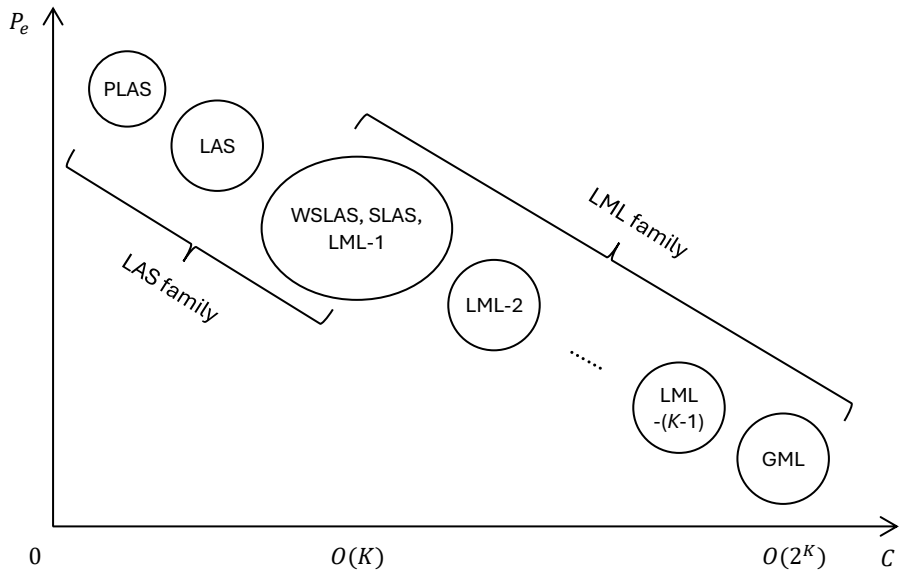

**Figure 2.** The relationships of the LML and LAS detectors in terms of error probability $P_e$ and computational complexity $C$.

### 3.6. Upper bounds on BER and lower bound on AME

Like other nonlinear detectors, the BER performance of the LAS detectors is difficult to analyze for finite-size MIMO channels. However, an upper bound of LAS BER can be obtained [19], which is comparable to that of the GML detector [18]. Let $F_k$ be the set of indecomposable error vectors affecting bit $k$ and $w(\boldsymbol{\varepsilon})$ be the weight of the error vector $\boldsymbol{\varepsilon} = 0.5(\boldsymbol{b} - \boldsymbol{b}^*)$ where $\boldsymbol{b}$ is the transmitted vector and $\boldsymbol{b}^*$ is the erroneous vector. Let $Q(x) = (2\pi)^{-0.5} \int_x^\infty \exp(-t^2/2)dt$. The following theorem is proved in [19].

*Theorem 4*: Given any initial detector $\boldsymbol{b}(0)$, the BER of the $k$th bit for the LAS detector associated with a sequence of $L(n)$ for $n \geq 0$ is upper bounded by

$$P_k^{\text{LAS}}(\sigma) \leq \sum_{\boldsymbol{\varepsilon} \in F_k} 2^{-w(\boldsymbol{\varepsilon})} Q\left(\frac{\boldsymbol{\varepsilon}^T(2\mathbf{H} - \mathbf{T})\boldsymbol{\varepsilon}}{\sigma\sqrt{\boldsymbol{\varepsilon}^T \boldsymbol{\varepsilon}}}\right) \tag{34}$$

where $\mathbf{T} = \text{diag}(t_1^*, \ldots, t_K^*)$ with $t_k^*$ given in Proposition 1.

Confirming the observation on the fixed-point region in Eq. (27), a decrease of the thresholds can decrease the upper bound. Therefore, three changes can result in a decrease in the upper bound: (i) a decrease in the total number of the update candidates, (ii) a decrease in the absolute values of the crosscorrelations between the update candidates, and (iii) a decrease in the signal powers of the update candidates.

In particular, by letting $t_k^* = A_k^2$ in Eq. (34), the BER upper bound for all the WSLAS/ LML-1 detectors is obtained as in the following corollary.

*Corollary 1*: The BER of the $k$th bit for the WSLAS/LML-1 detectors is upper bounded by

$$P_k^{\text{LML-1}}(\sigma) \leq \sum_{\boldsymbol{\varepsilon} \in F_k} 2^{-w(\boldsymbol{\varepsilon})} Q\left(\frac{\boldsymbol{\varepsilon}^T(2\mathbf{H} - \mathbf{A}^2)\boldsymbol{\varepsilon}}{\sigma\sqrt{\boldsymbol{\varepsilon}^T\boldsymbol{\varepsilon}}}\right). \tag{35}$$

It is obvious that the WSLAS/LML-1 detectors achieve the least upper bound in the family of LAS detectors. Since any LML-$J$ vector is also an LML-1 vector in terms of Eqs. (30) and (31), the upper bound in Eq. (35) applies to all the LML-$J$ detectors for any $J$.

The BER upper bound in Eq. (34) of the LAS detector is comparable with the upper bound of the GML detector obtained by Verdú [18]

$$P_k^{\text{GML}}(\sigma) \leq \sum_{\boldsymbol{\varepsilon} \in F_k} 2^{-w(\boldsymbol{\varepsilon})} Q\left(\frac{\sqrt{\boldsymbol{\varepsilon}^T\mathbf{H}\boldsymbol{\varepsilon}}}{\sigma}\right). \tag{36}$$

Note that $\boldsymbol{\varepsilon}^T(2\mathbf{H} - \mathbf{T})\boldsymbol{\varepsilon}/\sqrt{\boldsymbol{\varepsilon}^T\boldsymbol{\varepsilon}} \leq \boldsymbol{\varepsilon}^T(2\mathbf{H} - \mathbf{A}^2)\boldsymbol{\varepsilon}/\sqrt{\boldsymbol{\varepsilon}^T\boldsymbol{\varepsilon}} \leq \sqrt{\boldsymbol{\varepsilon}^T\mathbf{H}\boldsymbol{\varepsilon}}$. The second equality is true iff $w(\boldsymbol{\varepsilon}) = 1$. That is, the LML-1 detector is the GML detector when a transmitted vector $\boldsymbol{b}$ and an erroneous $\boldsymbol{b}^*$ differ by one bit. The upper bounds in Eqs. (34)-(36) indicate the BER difference of the LAS, LML-1, and GML detectors.

For a detector that achieves BER $P_k(A)$ for the $k$th bit with energy $A^2$, the multi-bit (or multiuser) efficiency $\eta = Q^{-2}[P_k(A)]\sigma^2/A^2$ measures the efficiency of signal power usage by the detector in comparison with the MF detector in the single-bit channel with bit energy $A^2$. The asymptotic multi-bit efficiency (AME) is the limit of $\eta$ as the noise vanishes $\sigma \to 0$, and thus measures the efficiency of signal power usage in the high SNR regime [18]. By Eq. (34), the following corollary is obtained [19].

*Corollary 2*: The AME of the LAS detector is lower bounded by

$$\eta_k^{\text{LAS}}(\sigma) \geq \min_{\boldsymbol{\varepsilon} \in F_k}^2 \left\{\frac{[\boldsymbol{\varepsilon}^T(2\mathbf{H} - \mathbf{T})\boldsymbol{\varepsilon}]^+}{A_k\sqrt{\boldsymbol{\varepsilon}^T\boldsymbol{\varepsilon}}}\right\}; \tag{37}$$

particularly the AME for all the WSLAS/LML-1 detectors is lower bounded by

$$\eta_k^{\text{LAS}}(\sigma) \geq \min_{\boldsymbol{\varepsilon} \in F_k}^2 \left\{\frac{[\boldsymbol{\varepsilon}^T(2\mathbf{H} - \mathbf{A}^2)\boldsymbol{\varepsilon}]^+}{A_k\sqrt{\boldsymbol{\varepsilon}^T\boldsymbol{\varepsilon}}}\right\} \tag{38}$$

where $[z]^+ = \max\{0, z\}$.

Unit AME is particularly interesting. In the high SNR regime, BER is dominated by the minimum distance from the transmitted signal to the decision regions of error signals. If the AME for bit $k$ is unit, the minimum distance is determined by the single error and then bit $k$'s BER achieves the single bit bound $Q(A_k/\sigma)$ as if there was no interfering bit. It follows from Eq. (36) that the GML detector achieves unit AME in the channels such that [18]

$$d^{\text{GML}}(\boldsymbol{\varepsilon}) \equiv \sqrt{\boldsymbol{\varepsilon}^T\mathbf{H}\boldsymbol{\varepsilon}} \geq A_k, \qquad \forall \boldsymbol{\varepsilon} \in F_k. \tag{39}$$

However, none of the well-known suboptimum detectors is known to achieve unit AME except in some trivial two-bit channels and the orthogonal $K$-bit channel [18]. In contrast, like the GML detector, the LAS detector achieves the following result [19].

*Corollary 3*: If for each $\boldsymbol{\varepsilon} \in F_k$ with each $w(\boldsymbol{\varepsilon}) = m \leq K$ where $K$ can be finite or tend to infinity,

$$d^{\text{LAS}}(\boldsymbol{\varepsilon}) \equiv \frac{\boldsymbol{\varepsilon}^T(2\mathbf{H}-\mathbf{T})\boldsymbol{\varepsilon}}{\sqrt{\boldsymbol{\varepsilon}^T\boldsymbol{\varepsilon}}} \geq A_k, \tag{40}$$

then the LAS detector achieves unit AME for bit $k$. Furthermore, if Eq. (40) holds for all bits, then the LAS detector achieves unit AME for all bits. In particular, these are all true for the WSLAS/LML-1 detectors if

$$d^{\text{LML}-1}(\boldsymbol{\varepsilon}) \equiv \frac{\boldsymbol{\varepsilon}^T(2\mathbf{H}-\mathbf{A}^2)\boldsymbol{\varepsilon}}{\sqrt{\boldsymbol{\varepsilon}^T\boldsymbol{\varepsilon}}} \geq A_k. \tag{41}$$

The performance of GML, LAS, and LML-1 detectors is determined by their signal distance $d^{\text{GML}}(\boldsymbol{\varepsilon})$, $d^{\text{LAS}}(\boldsymbol{\varepsilon})$, and $d^{\text{LML}-1}(\boldsymbol{\varepsilon})$. The larger the signal distance, the lower the error probability. For the channels that satisfy the condition of Eq. (40) (or (41)), the BER of the LAS (or WSLAS/ LML-1) detector in the high SNR region is dominated by the single-bit error signals and approaches the single-bit bound asymptotically. There exist many channels that satisfy Eqs. (40) and (41) regardless of $K$ [19].

## 4. The WSLAS/LML-1 Detectors Approaching Optimum Performance in Large MIMO Channels

Among all the detectors in the LML and LAS families, the WSLAS/LML-1 detector is most interesting. First, with a per-bit computational complexity linear in $K$, an LML-1 detector is the simplest among all LML-$J$ detectors for any neighborhood size $J$. Second, an LML-1 detector can be realized by a WSLAS detector with efficient computations [11-14, 21-25]. Third, a WSLAS/LML-1 detector is the optimum detector in the LAS family. Finally, as presented below, in large MIMO channels, the WSLAS/LML-1 detectors can approach the GML performance in a wide range of channel load and SNR [21-25].

In a general large MIMO channel, $K \to \infty$ and $N \to \infty$ and the channel load $\alpha = K/N$ is fixed in the large-system limit. The $k$th channel vector is $\boldsymbol{s}_k = (s_{1k}, \ldots, s_{Nk})^T/N^{0.5}$ where $s_{ik}$'s are i.i.d. with zero mean and unit variance. Therefore, in the large-system limit, the channel vectors converge to the unit length, and the crosscorrelation between two channel vectors converges to zero almost surely, that is, for $j \neq k$

$$\lim_{N \to \infty} \|\boldsymbol{s}_k\| = 1, \quad \lim_{N \to \infty} \boldsymbol{s}_j^H \boldsymbol{s}_k = 0, \quad \text{a.s.} \tag{42}$$

where $H$ is replaced by $T$ as the BPSK signal is considered throughout. The empirical distribution of symbol energies $A_k^2$ converges to a distribution function. The ratio of neighborhood size $J$ to the number of bits $K$ is denoted by $\beta = J/K$ where $J$ can also grow with $K$. As the system size tends to infinity, $\beta \to 0$ for any LML-$J$ detector for a fixed $J$, and $\beta = 1$ for the GML detector.

The general large MIMO channel model with the property of Eq. (42) includes the massive antenna MIMO channels [27], the large random spreading (LRS) CDMA channels with long and short sequences with/without spread spectrum and/or time [21, 22], the LRS CDMA channels with sparse sequences [23, 24], and any large MIMO channel that employs a combination of the multiantenna, CDMA, spectral-temporal spreading sequences, and sparse sequence techniques.

The most important property of the large MIMO channel is Eq. (42). As the channel dimension increases $N \to \infty$, the crosscorrelation matrix $\mathbf{R}$ becomes more and more like an identity matrix $\mathbf{I}$, and the channel becomes more and more like an orthogonal channel. Consequently, the overlapped LML region of multiple vectors in Eq. (12) becomes smaller and smaller, the LML vectors in Eq. (9) become fewer and fewer, and the error probability of an LML detector becomes closer and closer to that of the GML detector. In the large-system limit as $N \to \infty$, the large MIMO channel presents several properties as follows.

*4.1. LML characteristic of large MIMO channels*

The LML characteristic is one of the most important properties of the large MIMO channels. It is proved for the LRS CDMA channels in [21, 22] and here is extended to the general large MIMO channels as the theorem below. Let $E$ denote the set of error vectors $\boldsymbol{\varepsilon}$'s, and $I(\boldsymbol{\varepsilon})$ and $w(\boldsymbol{\varepsilon})$ the index set of nonzero elements and the weight of the error vector $\boldsymbol{\varepsilon}$, respectively.

*Theorem 5:* In the large MIMO channel with any $\alpha > 0$, (i) given any positive integers $M_1 < M_2$, $d^{\text{GML}}(\boldsymbol{\varepsilon}_1) < d^{\text{GML}}(\boldsymbol{\varepsilon}_2)$ a.s. for any $\boldsymbol{\varepsilon}_1, \boldsymbol{\varepsilon}_2 \in E$ such that $I(\boldsymbol{\varepsilon}_1) \subset I(\boldsymbol{\varepsilon}_2)$ and $w(\boldsymbol{\varepsilon}_1) \leq M_1 < w(\boldsymbol{\varepsilon}_2) \leq M_2$; (ii) for any $\delta > 0$, there exists $M \geq 2$ such that $d^{\text{GML}}(\boldsymbol{\varepsilon}) > \delta$ a.s. for any $\boldsymbol{\varepsilon} \in E$ with $w(\varepsilon) \geq M$.

As indicated in [21, 22], the LML characteristic means that in the large MIMO channel limit, "the mapping of **SA** from $\Omega_K$ to $\{\mathbf{SA}\boldsymbol{b}: \boldsymbol{b} \in \Omega_K\} \subset \mathbb{R}^N$ retain the local topology of $\Omega_K$ at any bit vector. Standing at the transmitted signal in the $\boldsymbol{r}$ space, one would typically see that the error signals with larger error weights are farther away, and all the error signals with the error weights tending to infinity are infinitely far away. Since the GML decision is based on the nearest distance from $\boldsymbol{r}$ to a signal $\mathbf{SA}\boldsymbol{b}$, the GML BER in the high SNR regime is dominated by the signals that have a one-bit error. This suggests that to achieve the GML detection, one would, without the exhaustive search over the entire set $\Omega_K$, perform only an LML detection."

*4.2. Achievability of GML performance by WSLAS/LML-1 detectors*

The LML characteristic of large MIMO channels enables the WSLAS/LML-1 detectors to achieve the GML performance in a range of channel load and SNR. The following theorems are proved for LRS CDMA channels [21, 22], and here are extended to the general large MIMO channels. The critical channel load is equal to $\alpha^* \equiv 1/2 - 1/(4\ln 2)$.

*Theorem 6:* In the large MIMO channel with $\alpha < \alpha^*$, the AMEs of all the LML detectors converge a.s. to one.

In the large MIMO channel with $\alpha < \alpha^*$, Theorem 6 indicates that the AMEs of the WSLAS/LML-1 detectors converge to one and, therefore, their BERs approach to the single-bit BER bound in the high SNR regime.

*Theorem 7:* In the large MIMO channel where $\alpha < \alpha^*$ and $\sqrt{N}\sigma = c \in (0, \infty)$ fixed, an LML-1 point is a.s. the GML point.

As pointed out in [21, 22], in the condition of Theorem 7, "there is typically no LML but the GML point. The likelihood function is sufficiently smooth. Thus, the linear-complex WSLAS detector (and other detectors) with local search can reach the GML point that is usually NP-hard to obtain. All the results are also applicable to the LML detectors with neighborhood sizes greater than one [15-17]."

Although the results in Theorem 6 and Theorem 7 hold with channel load $\alpha < \alpha^* \cong 0.1393$, the simulation results demonstrate that in random spreading CDMA and equal bit powers, the WSLAS detector with a linear per-bit complexity approaches the GML BER in all SNR with channel load as high as 1.05 bits/s/Hz where the dimension is second times Hz [21, 22].

*4.3. BER of WSLAS/LML-1 detectors*

The BER of the GML detector in the LRS CDMA channel is obtained by the replica analysis by Tanaka [61] and Guo and Verdú [62]. Similarly, we obtain [25] the BER of the WSLAS/LML-1 detectors in the general massive MIMO channels by the replica analysis and by applying the method of Gaussian approximation to the boundary of the fixed-point regions in Eq. (29). The results are presented in the propositions below and can be presented in theorems when the replica method is mathematically proved.

*Proposition 2:* In the large MIMO channel, for a bit with energy $A^2$, the BER of WSLAS/LML-1 detectors is the solution of the fixed-point equation

$$p_e(A) = Q\left(\frac{A}{\sqrt{\sigma^2 + 4\alpha \mathbf{E}_A(A^2 p_e(A))}}\right) \quad (43)$$

where $\mathbf{E}_A$ is the expectation with respect to $A$. $p_e(A)$ is also one of the BER solutions of the GML detector and all LML-$J$ detectors for $J \geq 2$.

Proposition 2 points out that the WSLA/LML-1 detectors with linear per-bit complexity can achieve the BER of the GML detector with NP-hard complexity in the large MIMO channels. It also explains why the family of LAS detectors and other detectors with local search can approach the GML performance in the massive antenna MIMO channels [26-58].

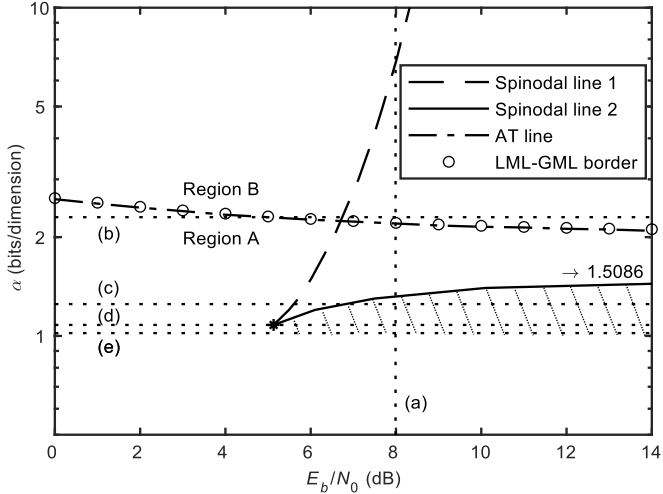

**Figure 3.** A practical region of $\alpha$ and $E_b/N_0$. Lines (a) – (e) indicate $\alpha$ and $E_b/N_0$ used in Figures 4 (a) – (e), respectively. The spinodal lines 1, 2 intersect at *: $(\alpha, E_b/N_0) = (1.08, 5.13)$ (bits/dimension, dB).

As shown in Figures 3 and 4 [25], the LML-1 BER in Eq. (43) is numerically evaluated and compared with the GML BER [61, 62] in an equal-energy distribution of $A_k = A, \forall k$, under various channel loads $\alpha$ and SNR $E_b/N_0$. The BER of a SLAS detector [21, 22] in simulation is also obtained with a random initial vector. The single-bit bound (SBB) on BER is also shown. Several results can be observed in the figures.

First, like the GML BER [61], as $\alpha$ becomes larger, multiple LML-BERs coexist.

Second, as shown in Figure 3, the AT line divides the region of channel load and SNR $(\alpha, E_b/N_0)$ into two regions. In region A, the LML-1 and GML detectors perform identically with the only saddle-point solution of $q = 1$. In region B, the GML detector has two bad solutions with $q < 0.7$ and $q = 1$, respectively. The LML-1 detector performs worse than the GML detector only for the additional bad solution of the high BER with $q < 0.7$, which the LML-1 detector cannot achieve.

Third, it is particularly practically interesting that there is a region below the spinodal line 2 denoted by the sloped dotted lines in Figure 3. In this region, the LML-1 BER is identical to the GML BER, and the channel load is greater than 1 bit/dimension. Specifically, for $\alpha \in [1, 1.08)$ bit/dimension, the SNR can be any $E_b/N_0 \in (-\infty, \infty)$. On the other hand, in the high SNR region as $E_b/N_0 \to \infty$, the channel load can be as high as $\alpha \cong 1.5086$ bits/dimension.

Fourth, for $\alpha \in [1, 1.08)$ bit/dimension, the practical SLAS detector (an LML-1 detector) in simulation can achieve the GML/LML-1 BER in the large-system limit. It implies that the linear-complex WSLAS detector can approach the GML performance in massive antenna MIMO systems. More discussions of Figure 4 are presented in Section 4.5.

In summary, in large MIMO channels, the linear-complex WSLAS/LML-1 detectors can achieve all the good BER solutions of the NP-hard GML detector.

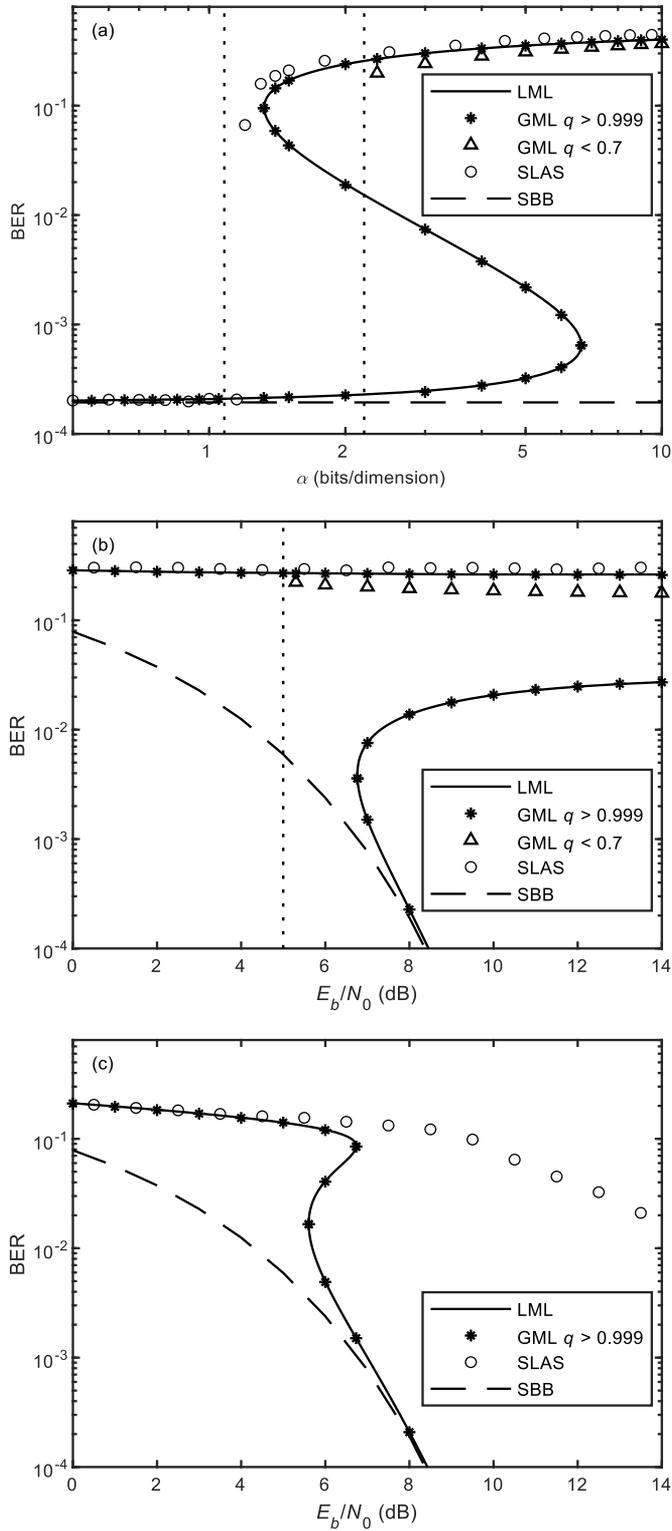

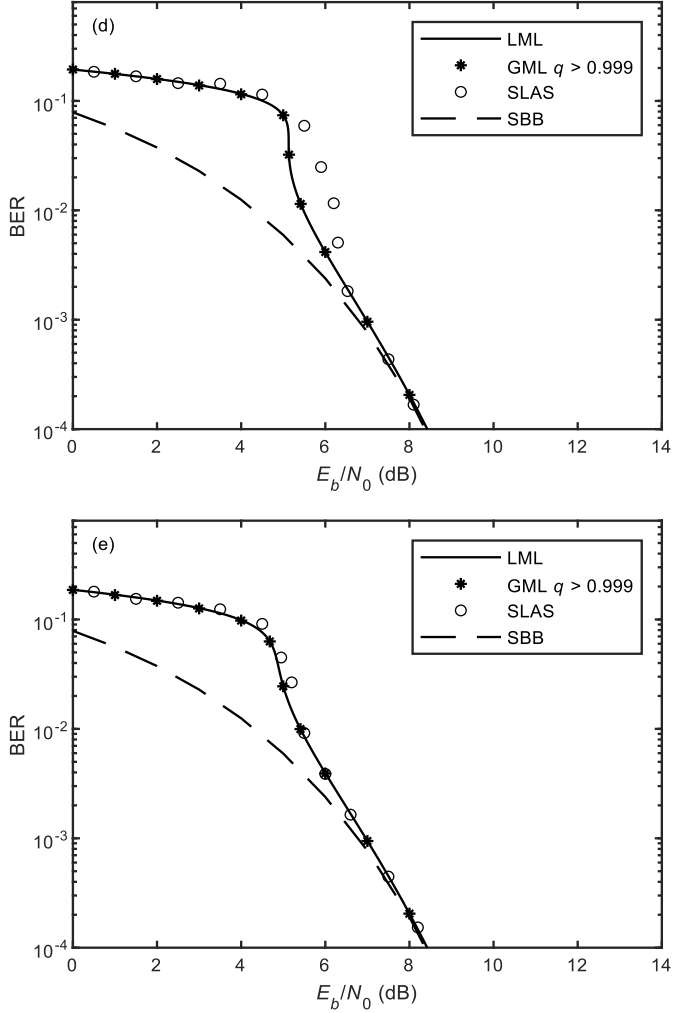

**Figure 4.** (a) BER versus $\alpha$ with $E_b/N_0 = 7.99$ dB. The left line is at $\alpha = 1.08$ bits/dimension, and the right at $\alpha = 2.20$ bits/dimension which separates regions A and B. BER versus $E_b/N_0$. (b) BER versus $E_b/N_0$ with $\alpha = 2.30$ bits/dimension. The vertical line indicates $E_b/N_0 = 5.0$ dB that separates regions A and B. (c) BER versus $E_b/N_0$ with $\alpha = 1.25$ bits/dimension. (d) BER versus $E_b/N_0$ with $\alpha = 1.08$ bits/dimension. (e) BER versus $E_b/N_0$ with $\alpha = 1.02$ bits/dimension. The values of $(\alpha, E_b/N_0)$ in (c) – (e) all are in region A; therefore, the gap between the LML and GML BERs is small, and so is the gap between their multiuser efficiencies.

### 4.4. Achievable channel load

By means of Eq. (43), the multi-bit efficiency of the LML-1 detector is the solution to the following fixed-point equation $\eta = \sigma^2/\{\sigma^2 + 4\alpha \mathbf{E}_A[A^2 p_e(A)]\}$ and the AME is

$$\hat{\eta} = \lim_{\sigma \to 0} \frac{\sigma^2}{\sigma^2 + 4\alpha \mathbf{E}_A[A^2 p_e(A)]}. \tag{44}$$

Since $p_e(A)$ can be multiple-valued, and so are the multi-bit efficiency and AME. In the large-system limit, the BER for the bit of energy $A^2$ is $p_e(A)$. If the AME equals one, in the high SNR regime, the LML-1 BER $p_e(A)$ can approach the SBB $p_s(A) = Q(A/\sigma)$ as if there was no interference bit.

The cutoff channel load (CCL) in the high SNR regime is defined as

$$\alpha^* = \sup_{\alpha > 0}\{\alpha: \hat{\eta} = 1\}.$$

The CCL $\alpha^*$ is the marginal channel load at which the bad solution disappears in the high SNR regime. All channel loads $\alpha < \alpha^*$ are asymptotically achievable, under which all bits can achieve the SBB BER in the high SNR regime. On the other hand, all channel loads $\alpha > \alpha^*$ are not achievable.

To obtain the CCL for the LML-1 detector in an equal-energy distribution, we have the following lemma [25].

*Lemma 1*: For the fixed-point equation of $p$

$$p = Q\left(\frac{1}{\sqrt{4\alpha p}}\right), \quad \alpha > 0, \tag{45}$$

there exist $\alpha_0 \cong 1.5086$ and $p_0 \cong 0.1169$ such that (i) if $\alpha < \alpha_0$, only $p = 0$ solves Eq. (45); (ii) if $\alpha = \alpha_0$, then both $p = p_0$ and $p = 0$ solves Eq. (45); (iii) if $\alpha > \alpha_0$, in addition to $p = 0$ there exist two points $p_1 \in (0, p_0)$ and $p_2 \in (p_0, 1)$ that solve Eq. (45).

By Eq. (43), as $\sigma \to 0$, the LML-1 BER in the equal-energy distribution is a fixed point that solves Eq. (45). In terms of Lemma 1, the LML-1 CCL is $\alpha^* = \alpha_0 \cong 1.5086$ bits/dimension. $\alpha_0$ is the marginal channel load such that the bad solution disappears in the high SNR regime. Consequently, the LML-1 and GML detectors have the same CCL.

It is interesting that the LML-1/GML CCL $\alpha_0$ is greater than 1 bit/dimension. In contrast, the channel load equals 1 bit/dimension for a set of orthogonal channel vectors and for the TDMA/FDMA systems where both transmitter and receiver have a single antenna. Moreover, the AMEs of the MMSE, decorrelating and MF detectors all are smaller than one [63] and, therefore, their CCLs are equal to zero.

The LML-1/GML CCL is low in the equal-energy distribution and can be arbitrarily high in an unequal-energy distribution. To obtain this, the result of LML-1 CCL is extended to an arbitrary energy distribution [25].

*Proposition 3*: The LML-1 CCL in an arbitrary energy distribution is equal to

$$\alpha^* = \frac{I}{4\mathbf{E}_A[A^2 Q(A/I^{1/2})]} \tag{46}$$

where the interference energy at $\alpha^*$ satisfies

$$I = \frac{\mathbf{E}_A^2[A^3 \exp(-A^2/(2I))]}{8\pi \mathbf{E}_A^2[A^2 Q(A/I^{1/2})]}. \tag{47}$$

To obtain $\alpha^*$ numerically in an arbitrary energy distribution, we can first iteratively solve $I$ from Eq. (47) and then obtain $\alpha^*$ from Eq. (46). Letting $A = 1$, Eqs. (46) and (47) can be applied to determine iteratively the LML-1 CCL $\alpha_0$ in the equal-energy distribution as given in Lemma 1.

As shown in Figure 5 for a two-class system, an unequal energy distribution can significantly increase the region of achievable channel load. As the energy difference between the two classes increases, the CCL monotonically increases and can be arbitrarily close to $\alpha_0/\lambda_1$ if $A_2$ is sufficiently small. This is true in general; that is, without limitation on the energy distribution, any channel load $\alpha > 0$ is achievable for the LML-1 detector. Moreover, $\alpha^* \geq \alpha_0$ where the equality holds iff all bits have equal energy. Hence, an unequal energy distribution is preferable to an equal energy distribution.

According to Eq. (43), the LML-1 detector has the total effective multi-bit interference $4\alpha\mathbf{E}_A[A^2 p_e(A)]$. A bit of energy $A^2$ interferes with others only through $4A^2 p_e(A)$. Below the spinodal line 2 of Figure 3 where the bad solution disappears, the BER $p_e(A)$ decreases exponentially fast to zero as the energy increases; therefore, a strong bit yields a weak interference. In particular, a bit with $A \to \infty$ does not interfere with others since $A^2 p_e(A) \to 0$ and thus the LML-1 detector is effective in near-far resistance [18]. Moreover, a weak bit also yields a weak interference. Consequently, the overall interference is low. This is why the LML-1 detector is superior to all linear detectors. For example, the linear MMSE detector has the multi-bit efficiency $\eta = \sigma^2/\{\sigma^2 + \alpha\mathbf{E}_A[A^2/(1 + \eta A^2 \sigma^{-2})]\}$ [63]. A bit of energy $A^2$ interferes others through $A^2/(1 + \eta A^2 \sigma^{-2})$, which does not vanish as energy increases; therefore, the overall interference is high.

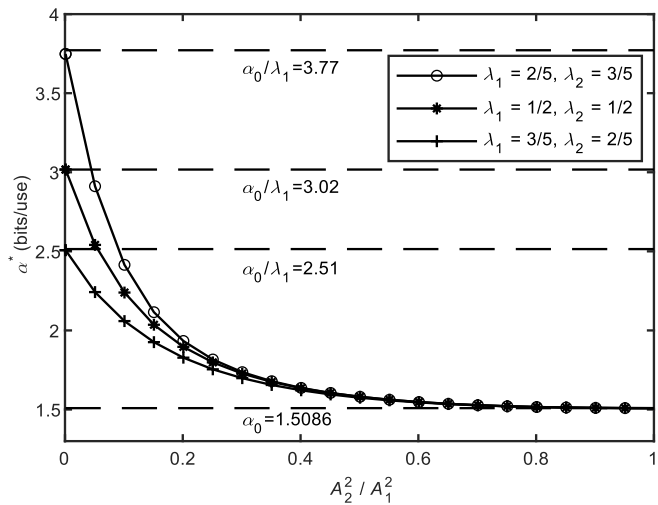

**Figure 5.** The cutoff channel load $\alpha^*$ versus energy distribution in the two-class system where the $i$th class has energy $A_i^2$ and percentage of population $\lambda_i$, $i = 1, 2$.

## 4.5. Practical achievability of GML BER

As pointed out in the preceding subsection, there is a special region of channel load and SNR as indicated by the sloped dotted lines in Figure 3. When the channel load $\alpha$ is lower than the spinodal line 2 where the bad solution disappears, there is only a good solution, and the LML-1 and GML detectors perform identically. Specifically, when $\alpha < 1.08$ bits/dimension, there is no other solution but a good solution for all SNR. If $E_b/N_0 > 5.13$ dB, the channel load can be higher than 1.08 bits/dimension. As SNR tends to infinity, the highest channel load is $\alpha = 1.5086$ bits/dimension. These facts imply the possible achievability of the NP-hard GML BER in a practical large MIMO system by a linear-complex LML-1 detector that uses only the local information of the current bit vector in a search.

To investigate the practicability, the SLAS detector [21, 22] is examined in the simulation. The SLAS detector, which belongs to the WSLAS detectors, is an LML-1 detector and has a linear per-bit computational complexity [21, 22]. In the simulation, the equal-energy system with $K = 8192$ bits is considered. A random bit vector is used for the SLAS detector to initiate the likelihood ascent search. The BER is obtained by averaging over a number of transmissions such that at least a total of 300 bit errors occur.

In Figure 4 (a) with $E_b/N_0 = 7.99$ dB. The SLAS detector behaves differently in three regimes of channel load. In the regime of $\alpha < 1.15$ bits/dimension where there is only the good LML-1/GML solution, the SLAS detector approaches the BER that is close to the SBB. In the regime between the spinodal line 2 and the AT line with $1.15 < \alpha < 2.20$ bits/dimension where one good and one bad LML-1/GML solutions coexist, the SLAS detector can only approach the bad BER. On the other hand, in the regime above the AT line (region B) with $\alpha > 2.20$ bits/dimension where the good LML-1/GML solution, the bad LML-1/GML solution, and the bad GML solution coexist, the SLAS detector can only approach the higher LML-1/GML BER of the bad solution. This confirms the preceding result that the lower GML BER of the bad solution with $q < 0.7$ cannot be approached by an LML-1 detector.

In Figure 4 (b) with $\alpha = 2.3$ bits/dimension, the left side of the vertical line with $E_b/N_0 = 5.0$ dB is in region A where there is only one bad LML-1/GML solution, and the right side is in region B where an additional bad GML solution appears.

However, as expected, in the entire considered region, the SLAS detector can only approach the bad LML-1/GML solution with the higher BER.

In Figure 4 (c) with $\alpha = 1.25$ bits/dimension, the system parameters $(\alpha, E_b/N_0)$ are in region A below the AT line, and therefore, the GML and LML-1 detectors perform identically. For $E_b/N_0 < 6.73$ dB above the spinodal line 2, the bad LML-1/GML solution exists, and the SLAS detector can only approach the bad solution. For $E_b/N_0 > 6.73$ dB below the spinodal line 2, only the good LML-1/GML solution exists. However, since $\alpha = 1.25$ bits/dimension is too close to the spinodal line 2, as shown in Figure 3, the SLAS detector approaches a BER with a big gap to the analytical solution. The reason is that when the system parameters $(\alpha, E_b/N_0)$ are close to the spinodal line 2, the SLAS performance is sensitive to the system size. It is observed that the BER gap decreases as the number of bits increases. That is, the number of bits $K = 8192$ is not sufficiently large for the SLAS detector to approach the analytical solution.

In Figure 4 (d) with $\alpha = 1.08$ bits/dimension that is farther away from the spinodal line 2, the SLAS detector can approach the good LML-1/GML BER but is slightly affected by the spinodal line 2 when $E_b/N_0$ is close to 5.13 dB at which the two spinodal lines intersect.

In Figure 4 (e) with $\alpha = 1.02$ bits/dimension where the system parameters $(\alpha, E_b/N_0)$ are far away from the spinodal line 2, the SLAS detector can approach the unique good LML-1/GML BER.

These simulation results demonstrate that the linear-complex SLAS detector using only the local information in the likelihood ascent search can approach the NP-hard GML performance in a practical large MIMO system. This is theoretically significant and practically useful when the LAS is rolled out and implemented on hardware suitable for fast parallel computations, as presented in the next section.

The numerical results demonstrate that in the large system limit, the BER difference between the LML with $\beta \to 0$ and the GML with $\beta = 1$ is small. The BER difference among all LML-$J$ detectors with any $\beta = J/K \in [0,1]$ must be small. This implies that in a sufficiently large practical MIMO system, the BER difference among all LML-$J$ detectors for $J > 1$ must be small. Since the SLAS/LML-1 detector in a system of $K = 8192$ bits can already approach the BER with a small gap to the analytical GML BER, any other LML-$J$ detector for $J > 1$ must not considerably improve the BER. On the other hand, the computational complexity of an LML-$J$ detector grows polynomially fast with $J$ [15-17]. Hence, in a practical, sufficiently large MIMO system, it is preferred to employ the WSLAS/LML-1 detector to achieve the best tradeoff between the BER performance and computational complexity.

## 5. Application to Next Generation Wireless Networks

The next generation 6G and beyond wireless network is expected to connect a large number of wireless devices and push the spectral load to its limit [59, 60]. On one hand, massive MIMO antennas can significantly increase the dimension of information transmission in a massive MIMO system. On the other hand, it is critical to exploit the spectral efficiency at the physical layer by an advanced detector. The family of LML detectors and the family of LAS detectors are particularly suitable for the symbol detector in a massive MIMO system to achieve optimum spectral efficiency. Among all the LML and LML detectors the WSLAS/LML-1 detector can achieve the GML detection while the per-bit complexity is linear and, therefore, is the best candidate for the 6G and beyond wireless networks. Aiming at this objective, several practical issues are addressed below.

## 5.1. Implementation of WSLAS detectors on a multilayer perception

Like all the LMLAS and LAS detectors, the WSLAS detector searches a bit vector in iteration. Iterative computation might be the major factor that hinders the WSLAS detector from being applied in practical MIMO systems where symbols must be detected in real time. To address the issue of iterations, we propose to roll out the iterations of the WSLAS detector.

Specifically, consider the neural network architecture of a multilayer perceptron. The search step $n$ indexes a layer. Each layer has $K$ neurons. The output state of the $k$th neuron in the $n$th layer is $b_k(n) \in \{-1,1\}$. Two adjacent layers are connected by the matrix $-\mathbf{H}$. According to Eq. (16), the $k$th neuron in the $n$th layer has the input of $-\sum_{i=1}^{K} H_{ki} b_i(n-1)$, the bias of $A_k y_k$, the neural threshold $t_k(n)$ calculated by Eq. (21), and by Eq. (20), the neural output function in Figure 6. The gradient in a layer is updated from the previous layer according to Eq. (18).

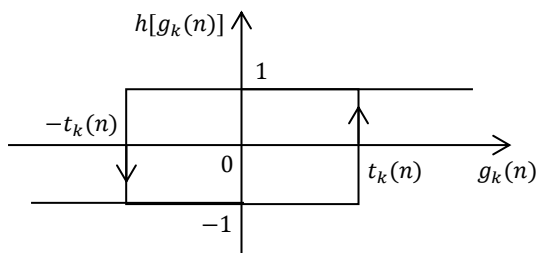

**Figure 6.** The output function of the $k$th neuron in the $n$th layer $b_k(n+1) = h[g_k(n)]$. The arrows indicate the directions of state transitions. The neural state does not change when $g_k(n) \in [-t_k(n), t_k(n)]$.

The WSLAS detector takes $n^*$ search steps to reach a fixed point, and $n^* = cK$ is linear in $K$, which is also the per-bit complexity in terms of the number of additions used per bit. Thus, the number of layers in the multilayer perceptron is equal to $cK$ where $c$ is much smaller than $K$. For example, the SLAS detector updates one bit in each step and $c$ is equal to the bit flip rate (BFR). In the simulations [21, 22], the number of bits is $1136 \leq K \leq 3328$, the channel load and SNR are in a wide practical range, and the MF detector is employed as the initial detector. The SLAS detector approaches the GML BER in the large system limit while the BFR is $0.01 < c < 0.49$. In the general case of $K = 3000$ and channel load $\alpha = 0.7$ bits/dimension, the BFR is $c = 0.2$ and then the number of layers is about $cK = 600$. In the worst case of $K = 3328$ and channel load $\alpha = 1.2$ bits/dimension, the BFR is $c = 0.49$ and the number of layers is about $cK = 1630$, a deep multilayer perceptron.

By implementing the WSLAS detector on the architecture of a multilayer perception, the receiver can demodulate the transmitted symbols one vector after another in the pipeline and in real time. The only cost of time incurred is the latency in the computation time over the $cK$ layers. Hence, the rollout of the WSLAS detector on a multilayer perceptron can be implemented in practical MIMO systems to achieve optimum performance in symbol detection and spectral efficiency.

## 5.2. EHE and FMD criteria

The latency of the WSLAS detector on a multilayer perceptron depends on $n^*$, the number of steps to reach a fixed point. To reduce the latency, three methods can be applied to decrease $n^*$ while guaranteeing convergence to an LML-1 point.

First, the search sequence $L(n), n \geq 0$ can be properly designed. In a few initial steps, more bits are updated in each step with a large $|L(n)|$ in order to converge fast in fewer steps. Gradually fewer bits are updated in each step with a smaller $|L(n)|$. Finally, one bit is updated in each step to converge to an LML-1 vector.

Second, the fastest metric descent (FMD) criterium [11] can be applied. In each step, only the bits with the largest $|g_k(n)| - 0.5A_k$ are updated. Under the FMD criterium, the WSLAS detector can converge to an LML-1 vector in fewer steps and, therefore, the number of layers can be smaller.

Third, the eliminating highest error (EHE) criterium [11] can be employed. In each step, only the bits with the highest $g_k(n)/A_k$ are updated. The higher the relative gradient component $g_k(n)/A_k$, the higher the probability of correct bit flip. Hence, the EHE criterium can enable the WSLAS detector to converge to a fixed point in fewer steps, and then the number of layers can be smaller.

*5.3. Random sparse codes*

Random sparse codes can be applied to transmitted bits. The $k$th bit is multiplied by code vector $\boldsymbol{c}_k = (c_{k1}, \ldots, c_{kM})^T$ before transmission. Each code vector $\boldsymbol{c}_k \in \{-L^{-0.5}, 0, L^{-0.5}\}^M$ has $L$ nonzero chips that are randomly located and equiprobably take on $\pm L^{-0.5}$. When $L/M$ is small, the codes are sparse as most of the chips are zero. The advantage of random sparse codes is that only the nonzero chips need transmission power and, therefore, the overall transmission power is much less.

In the simulation [23, 24], the number of bits is $16 \leq K \leq 2000$, SNR is $E_b/N_0 = 8$ dB, and channel load is $\alpha = 0.8$ bit/dimension. The results demonstrate that when there are as few as $L = 16$ nonzero chips, the WSLAS detector can already approach the GML/LML-1 BER with non-sparse random codes. This implies that the WSLAS detector with a set of random sparse codes can significantly reduce transmission power while achieving optimum performance in massive MIMO channels.

*5.4. LML-J for small MIMO channels*

In a small or middle MIMO channel, the numbers of Tx and Rx antennas are small. The likelihood function is rough and presents more LML points with low likelihood [21, 22]. In general, the WSLAS detector is more likely trapped in an LML-1 point with low likelihood. Consequently, the WSLAS/LML-1 detector often performs far worse than the GML detector. To address the problem, the EHE and FMD criteria [11] can be applied. In addition, other methods can also be considered. For example, a more complex and highly performed MMSE detector can be employed as the initial detector. Another method is to initial the WSLAS detector with multiple random vectors and select the optimum fixed point with the maximum likelihood as the demodulated vector.

In small or middle MIMO channels, the computational complexity of the WSLAS detector in the order of $O(K)$ is small. Then an LML-$J$ detector with $J \geq 2$ can be applied and the per-bit computational complexity is in the order of $O(\sum_{i=1}^{J}\binom{K}{i})$. For example, suppose the BFR is $c = 0.5$ in all cases. In a massive MIMO channel where $K_m = 10000$ bits are transmitted, the per-bit computational complexity is $cK_m = 5000$. In a small MIMO channel where $K_s = 100$ bits are transmitted, an LML-2 detector has the per-bit computational complexity about $c(K_s + 0.5K_s^2) = 2550$, in the same order as the WSLAS/LML-1 detector in the massive MIMO channel.

# 6. Conclusions

The family of local maximum likelihood (LML) detectors consists of all LML-$J$ detectors that each achieve the maximum likelihood in its neighborhood of size

$J$, including the global maximum likelihood (GML) detector that achieves the maximum likelihood in the entire set of bit vectors. The family of likelihood ascent search (LAS) detectors each searches out a sequence of vectors with a monotonical likelihood ascent and, therefore, converges to a fixed point with a finite number of steps. The wide-sense sequential LAS (WSLAS) detectors are particularly interesting. First, they are the LML-1 detectors and belong to both families. Second, their computational complexity per bit equals $cK$ linear in $K$ with $c < 0.5$. Third, they can approach the NP-hard GML detector in large MIMO channels. Fourth, under several criteria, they can converge to an LML-1 vector in fewer steps. Finally, the WSLAS detectors, like all LAS detectors, can be implemented on the architecture of a multilayer perceptron to achieve real-time symbol detection.

The general large MIMO channel is proven to possess the LML characteristic and, therefore, a local search detector with likelihood ascent, like the WSLAS detectors, can approach the GML detection. Moreover, in the large system limit, the LML-1 and GML detectors are proven to perform identically in a practically wide range of channel load and SNR, and the LML-1/GML BER achieves the single-bit BER in the high SNR regime with the channel load as high as up to 1.5086 bits/dimension.

With the capability of exploiting the LML characteristic and the potential high spectral efficiency of massive MIMO channels, the families of LAS and LML detectors are expected to play a role in the next generation 6G and beyond wireless networks.

## References

1. J. K. Paik and A. K. Katsaggelos, "Image restoration using a modified Hopfield network," *IEEE Trans. Image Proc.*, vol. 1, no. 1, pp. 49-63, 1992.
2. Y. Sun and S. Yu, "A modified Hopfield neural network used in bilevel image restoration and reconstruction," in *Proc. Int. Symp. Inform. Theory & its Appl.*, 1992.
3. Y. Sun and S. Yu, "An eliminating highest error criterion in Hopfield neural network for bilevel image restoration," in *Proc. Int. Symp. Inform. Theory & its Appl.*, 1992.
4. Y. Sun, J.-G. Li and S.-Y. Yu, "Improvement on performance of modified Hopfield neural network for image restoration," *IEEE Trans. Image Proc.*, vol. 4, no. 5, pp. 688-692, 1995.
5. Y. Sun, "A generalized updating rule for modified Hopfield neural network," in *Proc. Int. Conf. Neural Networks* (ICNN'97), 1997.
6. Y. Sun, "A generalized updating rule for modified Hopfield neural network for quadratic optimization," *Neurocomputing*, vol. 19, no. 1-3, pp. 133-143, 1998.
7. Y. Sun, "Hopfield neural network based algorithms for image restoration and reconstruction. I. Algorithms and simulations," *IEEE Trans. Signal Processing*, vol. 48, no. 7, pp. 2105-2118, 2000.
8. Y. Sun, "Hopfield neural network based algorithms for image restoration and reconstruction. II. Performance analysis," *IEEE Trans. on Signal Processing*, vol. 48, no. 7, pp. 2119-2131, 2000.
9. J. J. Hopfield and D. W. Tank, "'Neural' computation of decisions in optimization problems," *Biol. Cyb.*, vol. 52, no. 3, pp. 141-152, 1985.
10. Y.-T. Zhou, R. Chellappa, A. Vaid and B. K. Jenkins, "Image restoration using a neural network," *IEEE Trans. Acoust., Speech, Sig. Proc.*, vol. 36, no. 7, pp. 1141-1151, 1988.

11. Y. Sun, "Eliminating-highest-error and fastest-metric-descent criteria and iterative algorithms for bit-synchronous CDMA multiuser detection," in *Proc. IEEE Int. Conf. Commu.,* ICC'98, 1998.
12. Y. Sun, "A generalized search rule of likelihood ascent search detectors for CDMA multiuser detection," in *Proc. 5th Conf. Info. Sys. Analy. & Synth./3rd Conf. Sys., Cybern. & Info.*, ISAS´99/SCI'99, 1999.
13. Y. Sun, "A family of linear complexity likelihood ascent search multiuser detectors for CDMA communications," in *Conf. Record 34th Asilomar Conf. Signals, Systems & Computers*, 2000.
14. Y. Sun, "A family of linear complexity likelihood ascent search detectors for CDMA multiuser detection," *in 2000 IEEE 6th Int. Symp. Spread Spectrum Tech. & Appl.*, ISSTA 2000, 2000.
15. Y. Sun, "A family of likelihood ascent search detectors achieving local maximum likelihood with an arbitrary neighborhood size for CDMA multiuser detection," in *Proc. 38th Ann. Allerton Conf. Commun., control, & computing*, 2000.
16. Y. Sun, "Local maximum likelihood multiuser detection," in *Proc. 34th Ann. Conf. Info. Sci. & Sys.*, CISS'2001, 2001.
17. Y. Sun, "Local maximum likelihood multiuser detection for CDMA communications," in *Pro. Int. Conf. Info. Techn.: Coding & Computing*, 2001.
18. S. Verdú, *Multiuser Detection*, Cambridge University Press, New York, 1998.
19. Y. Sun, "A family of likelihood ascent search multiuser detectors: an upper bound of bit error rate and a lower bound of asymptotic multiuser efficiency," *IEEE Trans. on Commu.*, vol. 57, no. 6, pp. 1743-1752, June 2009.
20. Y. Sun, "Local maximum likelihood multiuser detection," *ARL Collab. Tech. Alliance Commu. & Networks - Progress Report 4Q01*, Dec. 2001.
21. Y. Sun, "A family of likelihood ascent search multiuser detectors: approach to single-user performance via quasi-large random sequence CDMA," *arXiv preprint*, arXiv:0711.3867, 2007.
22. Y. Sun, "A family of likelihood ascent search multiuser detectors: approaching optimum performance via random multicodes with linear complexity," *IEEE Trans. on Commu.*, vol. 57, no. 8, pp. 2215-2220, Aug. 2009.
23. Y. Sun, "Quasi-large sparse-sequence CDMA: Approach to single-user bound by linearly-complex LAS detectors," in *42nd Ann. Conf. Info. Sci. Sys.*, CISS 2008, 2008.
24. Y. Sun and J. Xiao, "Multicode sparse-sequence CDMA: approach to optimum performance by linearly complex WSLAS detectors," *Wireless Personal Commu.*, vol. 71, no. 2, pp. 1049-1056, 2013.
25. Y. Sun, L. Zheng, P. Zhu, and X. Wang, "On optimality of local maximum-likelihood detectors in large-scale MIMO channels," *IEEE Trans. Wireless Commun.*, vol. 15, no. 10, pp. 7074-7088, Nov. 2016.
26. K. V. Vardhan, S. K. Mohammed, A. Chockalingam and B. S. Rajan, "A low-complexity detector for large MIMO systems and multicarrier CDMA systems," *IEEE J. Select. Areas Commu.*, vol. 26, no. 3, pp. 473-485, 2008.
27. M. A. Albreem, M. Juntti and S. Shahabuddin, "Massive MIMO detection techniques: A survey," *IEEE Commu. Surveys & Tutorials*, vol. 21, no. 4, pp. 3109-3132, 2019.
28. S. K. Mohammed, A. Chockalingam and B. S. Rajan, "A low-complexity near-ML performance achieving algorithm for large MIMO detection," in *2008 IEEE Int. Symp. Infor. Theory*, 2008.
29. A. Chockalingam, "Low-complexity algorithms for large-MIMO detection," in *4th Int. Symp. Commu., Control & Signal Proc.* (ISCCSP), 2010.
30. J. D. L. Ducoing, N. Yi, Y. Ma and R. Tafazolli, "Using real constellations in fully-and over-loaded large MU-MIMO systems with simple detection," *IEEE Wireless Commu. Lett.*, vol. 5, no. 1, pp. 92-95, 2016.

31. W. Liu, Y. Zhang and M. Jin, "Lagrangian detection for generalized space-shift keying MIMO systems," *IEEE Trans. Veh. Tech.*, vol. 66, no. 9, pp. 8585-8589, 2017.
32. N. R. Challa and K. Bagadi, "Likelihood ascent search detection for coded massive MU-MIMO systems to mitigate IAI and MUI," *Radioelectr. & Commu. Sys.*, vol. 63, no. 5, pp. 223-234, 2020.
33. L. Li, W. Meng and C. Li, "Semidefinite further relaxation on likelihood ascent search detection algorithm for high‐order modulation in massive MIMO system," *IET Commu.*, vol. 11, no. 6, pp. 801-808, 2017.
34. M. Solanki and S. Gupta, "Robust conjugate-gradient based LAS detector for massive MIMO systems," *Int. J. Electr.*, vol. 109, no. 5, pp. 794-810, 2022.
35. J. Hu, S. Song and Z. Wang, "A novel low-complexity massive MIMO detector with near-optimum performance," *in IEEE Int. Symp.Circ. Sys.* (ISCAS), 2024.
36. N. Srinidhi, S. K. Mohammed, A. Chockalingam and B. S. Rajan, "Low-complexity near-ML decoding of large non-orthogonal STBCs using reactive tabu search," in *IEEE Int. Symp. Infor. Theory*, 2009.
37. T. Datta, N. Srinidhi, A. Chockalingam and B. S. Rajan, "Random-restart reactive tabu search algorithm for detection in large-MIMO systems," *IEEE Commu. Letters*, vol. 14, no. 12, pp. 1107-1109, 2010.
38. T. Datta, N. Srinidhi, A. Chockalingam and B. S. Rajan, "A hybrid RTS-BP algorithm for improved detection of large-MIMO M-QAM signals," in *Nat. Conf. Commu.* (NCC), 2011.
39. N. Srinidhi, T. Datta, A. Chockalingam and B. S. Rajan, "Layered tabu search algorithm for large-MIMO detection and a lower bound on ML performance," *IEEE Trans. Commu.*, vol. 59, no. 11, pp. 2955-2963, 2011.
40. M. Karthikeyan and D. Saraswady, "Low complexity layered tabu search detection in large MIMO systems," *AEU-Int. J. Electr. Commu.*, vol. 83, pp. 106-113, 2018.
41. N. T. Nguyen, K. Lee and H. Dai, "QR-decomposition-aided tabu search detection for large MIMO systems," *IEEE Trans. Veh. Tech.*, vol. 68, no. 5, pp. 4857-4870, 2019.
42. S. Chakraborty, N. B. Sinha and M. Mitra, "Low complexity, pairwise layered Tabu search for large scale MIMO detection," *Wireless Personal Commu.*, vol. 128, no. 3, pp. 1689-1713, 2023.
43. A. K. Sah and A. K. Chaturvedi, "An unconstrained likelihood ascent based detection algorithm for large MIMO systems," *IEEE Trans. Wireless Commu.*, vol. 16, no. 4, pp. 2262-2273, 2017.
44. A. K. Sah and A. K. Chaturvedi, "Sequential and global likelihood ascent search-based detection in large MIMO systems," *IEEE Trans. Commu.*, vol. 66, no. 2, pp. 713 - 725, 2018.
45. A. K. Sah and A. K. Chaturvedi, "Reduced neighborhood search algorithms for low complexity detection in MIMO systems," in *IEEE Global Commu. Conf.* (GLOBECOM), 2015.
46. P. Mann, A. K. Sah, R. Budhiraja and A. K. Chaturvedi, "Bit-level reduced neighborhood search for low-complexity detection in large MIMO systems," *IEEE Wireless Commu. Lett.*, vol. 7, no. 2, pp. 146-149, 2018.
47. P. Li and R. D. Murch, "Multiple output selection-LAS algorithm in large MIMO systems," *IEEE Comm. Lett.*, vol. 14, no. 5, pp. 399-401, 2010.
48. A. A. J. Pereira, and R. Sampaio-Neto, "A random-list based LAS algorithm for near-optimal detection in large-scale uplink multiuser MIMO systems," in *19th Int. ITG Workshop Smart Antennas*, 2016.
49. M. Chaudhary, N. K. Meena and R. S. Kshetrimayum, "Local search based near optimal low complexity detection for large MIMO System," in *IEEE Int. Conf. Adv. Net. & Telecom. Sys.* (ANTS), 2016.
50. I. Aravindan, A. Kumar, T. C. Snehith, A. Padmakumar and R. Ramanathan, "A performance study of MIMO detectors in the presence of channel estimation errors," in *Int. Conf. Commu, Infor. & Computing Tech.* (ICCICT), 2015.
51. I. Chihaoui, M. L. Ammari and P. Fortier, "Improved LAS detector for MIMO systems with imperfect channel state information," *IET Commu.*, vol. 13, no. 9, pp. 1297-1303, 2019.
52. I. Chihaoui and M. L. Ammari, "LAS detector with soft-output MMSE initialization under imperfect channel estimation and channel correlation," *Wireless Personal Commu.*, vol. 108, pp. 213-220, 2019.
53. S. K. Mohammed, A. Zaki, A. Chockalingam and B. S. Rajan, "High-rate space–time coded large-MIMO systems: Low-complexity detection and channel estimation," *IEEE J. Select. Topics Signal Proc.*, vol. 3, no. 6, pp. 958-974, 2009.
54. Z. Qin, J. Xu, X. Tao and X. Zhou, "Improved depth-first-search sphere decoding based on LAS for MIMO-OFDM systems," in *IEEE 82nd Veh. Tech. Conf.* (VTC2015-Fall), 2015.
55. I. Chihaoui and M. L. Ammari, "Suited architecture for massive MIMO detector based on antenna selection and LAS algorithm," in *Int. Symp. Signal, Image, Video & Commu.* (ISIVC), 2016.
56. B. Cerato and E. Viterbo, "Hardware implementation of a low-complexity detector for large MIMO," in *IEEE Int. Symp. Circuits & Systems*, 2009.
57. L. Li, H. Hou and W. Meng, "Convolutional-neural-network-based detection algorithm for uplink multiuser massive MIMO systems," *IEEE Access*, vol. 8, pp. 64250-64265, 2020.
58. A. Ullah, W. Choi, T. M. Berhane, Y. Sambo and M. A. Imran, "Soft-Output Deep LAS Detection for Coded MIMO Systems: A Learning-Aided LLR Approximation," *IEEE Trans. Veh. Tech.*, 2024.
59. Y. Wu, X. Gao, S. Zhou, W. Yang, Y. Polyanskiy and G. Caire, "Massive access for future wireless communication systems," *IEEE Wireless Commu.*, vol. 27, no. 4, pp. 148-156, 2000.
60. Z. Wei, F. Liu, C. Masouros, N. Su and A. P. Petropulu, "Toward multi-functional 6G wireless networks: integrating sensing, communication, and security," *IEEE Commu. Mag.*, vol. 60, no. 4, pp. 65-71, 2022.
61. T. Tanaka, "A statistical-mechanics approach to large-system analysis of CDMA multiuser detectors," *IEEE Trans. on Inform. Theory*, vol. 48, pp. 2888-2910, Nov. 2002.
62. D. Guo, and S. Verdú, "Randomly spread CDMA: Asymptotics via statistical physics," *IEEE Trans. on Inform. Theory*, vol. 51, no. 6, pp. 1983-2010, June 2005.
63. D. N. C. Tse and S. V. Hanly, "Linear multiuser receivers: effective interference, effective bandwidth and user capacity," *IEEE Trans. Inform. Theory*, vol. 45, pp. 641-657, Mar. 1999.
64. L. B. Nelson and H. V. Poor, "Iterative multiuser receivers for CDMA channels: an EM-based approach," *IEEE Trans. Commun.*, vol. 44, no. 12, pp. 1700–1710, Dec. 1996.
65. M. K. Varanasi and B. Aazhang, "Near-optimum detection in synchronous code-division multiple access systems," *IEEE Trans. Commun.*, vol. 39, pp. 725–736, May 1991.
66. B. Wu and Q. Wang, "New suboptimal multiuser detectors for synchronous CDMA systems," *IEEE Trans. Commun.*, vol. 44, no. 7, pp. 782–785, July 1996.
67. D. Raphaeli, "Suboptimal maximum-likelihood multiuser detection of synchronous CDMA on frequency-selective multipath channels," *IEEE Trans. Commun.*, vol. 48, no. 5, pp. 875–885, May 2000.